\documentclass[12pt]{article}

\usepackage{latexsym}
\usepackage{amsfonts}
\usepackage{amsmath}

\newtheorem{remark}{Remark}
\newtheorem{theorem}{Theorem}
\newtheorem{corollary}{Corollary}

\begin{document}

\title{Cartan Normal Conformal Connections from Pairs of 2$^{nd}$ Order
PDE's}
\author{Emanuel Gallo*, Carlos Kozameh*, \and Ezra T. Newman$^{\#}$,
Kiplin Perkins$^{\#}$ \\ \\
$^{\#}$Dept of Physics and Astronomy, \\
Univ. of Pittsburgh, Pgh.PA 15260\\ \\
$^{*}$FAMAF, Universidad Nacional de \\
C\'{o}rdoba, 5000 C\'{o}rdoba, Argentina}
\date{15 April 2004.}
\maketitle

\begin{abstract}
We explore the different geometric structures that can be
constructed from the class of pairs of 2nd order PDE's that
satisfy the condition of a vanishing generalized W\"{u}nschmann
invariant. This condition arises naturally from the requirement of
a vanishing torsion tensor. In particular, we find that from this
class of PDE's we can obtain all four-dimensional conformal
Lorentzian metrics as well as all Cartan normal conformal $O(4,2)$
connections.

To conclude, we briefly discuss how the conformal Einstein
equations can be imposed by further restricting our class of PDE's
to those satisfying additional differential conditions.
\end{abstract}

\pagenumbering{arabic}

\section{Introduction}

This work, which deals with general relativity from a
non-conventional perspective, has several independent objectives.

One of them is to further demonstrate and develop the rich
geometric structures that are buried in a large class of
differential equations. Some of these structures have been known
for a long time,\cite{T,T1,L} while other parts are new. Here we
will concentrate on the geometry associated with pairs of 2nd
order PDE's with two independent and one dependent variable. Via
the emerging structure, the discussion will then be narrowed down
to a large special class of equations that in what are referred to
as the generalized W\"{u}nschmann class. We will describe the
differential geometry that is induced by the differential
equations of this class on the four-dimensional solution space of
the PDE's. They include the existence of all Lorentzian conformal
geometries as well as Cartan normal conformal connections. Since
all Lorentzian conformal metrics can be constructed from
\textit{some} pair of 2$^{nd}$ order PDE's, this means, in
particular, that all metrics that are conformally related to
vacuum Einstein metrics are included in this context or
discussion.

However, in order to gain perspective on this work, we point out and
summarize similar work that was done on a considerably simpler problem,
namely the study of the class of all 3$^{rd}$ order ODE's. Cartan\cite
{C,C1,C2,C3} and Chern\cite{Ch} showed how to construct, from the
differential equation, a triad and its associated connection on the
three-dimensional solution space of a generic third order ordinary
differential equation of the form

\begin{equation}
u^{\prime \prime \prime }=F(u,u^{\prime },u^{\prime \prime },s).
\label{third-order}
\end{equation}

The starting problem was to study the equivalence class of
equations under the group of contact transformations on the
$(u,u^{\prime },s)$ space. One remarkable result that follows from
the equivalence problem is that the equivalence classes of
$3^{rd}$ order ODE's split into two major classes: those with a
vanishing W\"{u}nschmann invariant\cite{W} and those with a
non-vanishing invariant. The W\"{u}nschmann invariant, $I[F],$ a
differential expression involving $F$ and its derivatives in all
four variables, was discovered by early workers in the theory of
differential equations and extensively used by Chern\cite{Ch}.
More specifically, when a third order ODE satisfies $I[F]=0,$ one
shows that the solution space, i.e., the three-dimensional space
of constants of integration, $(x^{a})$, possesses, directly from
the differential equation, Eq.(\ref{third-order}), a conformal
Lorentzian metric with the level surfaces of the solutions
themselves, $u=z(x^{a},s),$ forming a one parameter family of null
surfaces. All members of the equivalence class, under contact
transformations, yield the same conformal metric. The converse
statement is also true; namely, given a three-dimensional
conformal Lorentzian space-time, from any complete solution
($u=U(x^{a},s)$ ) of the eikonal equation, $g^{ab}\partial
_{a}U\partial _{b}U=0,$ one can obtain a third order ODE (by
differentiating with respect to $s$ three times and eliminating
the $x^{a})$ all belonging to the same equivalence
class\cite{FKN}. It was then shown by two different
methods\cite{C,C1,C2,C3,Ch,etal} that from the 3$^{rd}$ order ODE
with vanishing W\"{u}nschmann invariant one could generate on the
solution space a Cartan normal conformal connection with group
$O(2,3)$. The present work generalizes these ideas from 3$^{rd}$
order ODE's to pairs of 2$^{nd}$ order PDE's.

A second objective of this work is to prepare the basic structures
needed to code the Einstein equations (or more accurately, the
conformal Einstein equations) into the formalism of pairs of
PDE's. We first point out that there are two uses to the term
`conformal Einstein equations'; in one case there are differential
equations solely for conformal classes of Lorentzian metrics such
that \textit{there exists }a conformal factor whose choice
converts the class into a single vacuum metric that is a solution
of the vacuum Einstein equations, in the other case, the
differential equations involve both the metric \textit{and the
needed }conformal factor. Our interest, at the present, is only in
the first case.

More than twenty years ago the Null Surface Formulation of GR\cite
{KN,FNN,etal,FCN3,SCT} was introduced as an alternative tool to
study GR and in particular to capture the conformal degrees of
freedom. The novelty of the approach consisted in using null
surfaces as the main variables for the theory. The surfaces
themselves were obtained as solutions of an integrable pair of
second order PDE's for sections of a line-bundle over the
two-sphere with a complex function $\Lambda $ playing an analog
role to the above $F$. (Here, we use $S$ instead of $\Lambda .$)
The four-dimensional solution space of these equations emerged as
the space-time itself. An explicit algebraic method for
constructing the conformal metric, on the solution space, was
derived provided a certain differential condition on $\Lambda$ was
imposed. This condition, referred to then as the metricity
condition, written as $M=M[\Lambda ],$ was essential for the
formulation of the NSF. At the time we were not aware of the
different geometric meanings of this newly discovered expression
(even in the non-vanishing case) as a generalized W\"{u}nschmann
invariant for the second order PDE's under consideration.
Explicitly, in terms of $\Lambda ,$ we were able to construct, in
addition to the conformal metric, objects like the Weyl tensor,
the Riemann tensor, the Levi-Civita connection, etc. Although it
was not very elegant to mix the non-conformally invariant
connections with null surfaces for the construction of these
objects, at that time it was not clear how to produce a completely
conformally invariant formulation of conformal GR.

In this work we try to fill this gap. Starting with the pair of
PDE's that define our main variables and without prior assumptions
of a space-time, we construct what is called a conformal Cartan
connection on the solution space of these equations. A tetrad
(that becomes null), defined from the PDE's, is introduced on the
solution space. The requirement of a vanishing torsion uniquely
fixes the associated connection and imposes a differential
condition on the class of considered PDE's for our main variable.
Further, it guarantees the existence of a metric on the solution
space. The formalism is conformally covariant by construction; the
non-trivial part of its curvature are the Weyl and generalized
Cotton-York tensors. The Ricci tensor is coded into the Cartan
normal conformal connection one-form.

The conformal Einstein equations, in this new language, are to be
differential conditions imposed on the pair of PDE's that define
the characteristic surfaces. They are to be found by imposing
conditions on the Cartan conformal Curvature; both a cubic
algebraic condition and a differential condition, that is
equivalent to the vanishing Bach tensor.

In section II, we describe certain preliminary ideas and review
earlier results. Our main results are presented in section III
where we show how, from the first Cartan structure equation, we
find a conformal connection with the restriction on the class of
considered PDE's to the so-called generalized W\"{u}nschmann
class. Many of the explicit detailed expressions and proofs are
relegated to Appendices due to their length. In section IV we
discuss the second Cartan structure equation and the Cartan
curvature tensors. For clarity, in section V we give a brief
synopsis of the earlier sections. Section VI unifies the earlier
material into a Cartan normal conformal connection. In the section
VII, we discuss the issues of obtaining the conformal Einstein
equations.

\section{Preliminaries}

On a 2-dimensional space with coordinates $(s,s^{*})$ we consider the
following PDE's

\begin{eqnarray}
Z_{ss} &=&S(Z,Z_{s},Z_{s^{*}},Z_{ss^{*}},s,s^{*}),  \label{pde1} \\
Z_{s^{*}s^{*}} &=&S^{*}(Z,Z_{s},Z_{s^{*}},Z_{ss^{*}},s,s^{*}),
\nonumber
\end{eqnarray}
where the subscripts denote partial derivatives and $Z$ is a real
function of $(s,s^{*})$. Though it would have been equally
possible to treat $ (s,s^{*})$ as a pair of real variables it
turns out to be more useful to consider them as a
complex-conjugate pair. In that case the second equation is simply
the complex-conjugate of the first equation. In the following,
($^{*}$) will denote the complex-conjugate. By assumption, the
functions $S$ and $S^{*}$ satisfy the integrability conditions.
Solutions, $Z=Z(s,s^{*}),$ are two-surfaces in the six-dimensional
space, $J^{6},$ with coordinates
\begin{equation}
(Z,W,W^{*},R,s,s^{*})\equiv (Z,Z_{s},Z_{s^{*}},Z_{ss^{*}},s,s^{*}).
\label{notation1}
\end{equation}

For an arbitrary function $H=H(Z,W,W^{*},R,s,s^{*})$, the
total-derivatives in the $s$ and $s^{*}$ are

\begin{eqnarray}
\frac{dH}{ds} &\equiv &DH\equiv H_{s}+WH_{Z}+SH_{W}+RH_{W^{*}}+TH_{R},
\label{defD} \\
\frac{dH}{ds^{*}} &\equiv &D^{*}H\equiv
H_{s^{*}}+W^{*}H_{Z}+RH_{W}+S^{*}H_{W^{*}}+T^{*}H_{R},
\label{defDstar}
\end{eqnarray}
where

\begin{eqnarray}
T &=&D^{*}S,  \label{t} \\
T^{*} &=&DS^{*}.  \nonumber
\end{eqnarray}

The $T$ and $T^{*}$ are explicit functions of
$(Z,W,W^{*},R,s,s^{*})$ that are obtained in the following way.
Letting $H=S^{*}$ in Eq.(\ref{defD}) and $H=S$ in
Eq.(\ref{defDstar}), we get two equations containing $T$ and
$T^{*}$ . From them, we find
\begin{eqnarray}
T &=& \frac{S_{s^{*}}+W^{*}S_{Z}+
RS_{W}+S^{*}S_{W^{*}}}{1-S_{R}S_{R}^{*}}
\label{T} \\
&&+\frac{S_{R}(S_{s}^{*}+WS_{Z}^{*}+SS_{W}^{*}+ RS_{W^{*}}^{*})}{
1-S_{R}S_{R}^{*}}.  \nonumber
\end{eqnarray}

Note that the $D$ and $D^{*}$ are actually the coordinate vectors
$e_{s}$ and $e_{s^{*}}$, respectively. Hence,

\begin{eqnarray}
e_{s} &\equiv &D=\frac{d}{ds}= \frac{\partial }{\partial
s}+W\frac{\partial }{
\partial Z}+ S\frac{\partial }{\partial W} +
R\frac{\partial }{\partial W^{*}} + T
\frac{\partial }{\partial R},  \label{es} \\
e_{s^{*}} &\equiv &D^{*}=\frac{d}{ds^{*}}=\frac{\partial
}{\partial s^{*}} +W^{*}\frac{\partial }{\partial
Z}+R\frac{\partial }{\partial W}+S^{*}\frac{
\partial }{\partial W^{*}}+T^{*}\frac{\partial }{\partial R}.
\nonumber
\end{eqnarray}

Often, for detailed calculations, the following identities are
very useful. For $H=H(Z,W,W^{*},R,s,s^{*})$ and $y\in
\{Z,W,W^{*},R,s,s^{*}\}$

\begin{eqnarray}
D(H_{y}) &=& (DH),_{y}-(S_{y}H_{W}+T_{y}H_{R}+\delta
_{W,y}H_{Z}+\delta
_{R,y}H_{W^{*}}),  \label{comm} \\
D^{*}(H_{y}) &=&(D^{*}H),_{y}-(S_{y}^{*}H_{W^{*}}+T_{y}^{*}H_{R}+
\delta _{W^{*},y}H_{Z}+\delta _{R,y}H_{W}),  \nonumber
\end{eqnarray}
where $\delta _{y^{\prime }},_{y}$ is the Kronecker symbol.

With these definitions of $D$ and $D^{*}$ the integrability
conditions of (\ref{pde1}) are
\begin{equation}
D^{2}S^{*}=D^{*2}S.  \label{integra}
\end{equation}
In addition, the functions $S$ and $S^{*}$ are assumed to satisfy
the weak inequality
\begin{equation}
1-S_{R}S_{R}^{*}>0.  \label{inequal}
\end{equation}
From this inequality and the Frobenius theorem one can
show\cite{FKN} that the solutions depend on four parameters,
namely $x^{a}$, defining the space-time manifold, $\frak{M}^{4},$
as the solution space of the PDE's. We can thus write

\begin{eqnarray}
Z &=&Z(x^{a},s,s^{*}),  \label{notation2} \\
W &=&W(x^{a},s,s^{*}),  \nonumber \\
W^{*} &=&W^{*}(x^{a},s,s^{*}),  \nonumber \\
R &=&R(x^{a},s,s^{*}).  \nonumber
\end{eqnarray}

\begin{remark}
The space $J^{6}$ is foliated by the integral curves of $D$ and
$D^{*}$ which are labeled by $x^{a}.$ The above relations can be
interpreted as $(s,s^{*})$ dependent coordinate transformation
between the $(Z,W,W^{*},R)$ and the $x^{a}$.
\end{remark}

The exterior derivatives of (\ref{notation2}),

\begin{eqnarray}
dZ &=&Z_{a}dx^{a}+Wds+W^{*}ds^{*},  \label{1-forms} \\
dW &=&W_{a}dx^{a}+Sds+Rds^{*},  \nonumber \\
dW^{*} &=&W_{a}^{*}dx^{a}+Rds+S^{*}ds^{*},  \nonumber \\
dR &=&R_{a}dx^{a}+Tds+T^{*}ds^{*},  \nonumber
\end{eqnarray}
can be re-written as the Pfaffian system of four one-forms

\begin{eqnarray}
\beta ^{0}\, &\equiv &dZ-Wds-W^{*}ds^{*}=Z,_{a}dx^{a},
\label{betas} \\
\beta ^{+}\, &\equiv &dW-Sds-Rds^{*}=W,_{a}dx^{a},  \nonumber \\
\beta ^{-}\, &\equiv &dW^{*}-Rds-S^{*}ds^{*}=W^{*},_{a}dx^{a},
\nonumber \\
\beta ^{1}\, &\equiv &dZ-Tds-T^{*}ds^{*}=R,_{a}dx^{a}.  \nonumber
\end{eqnarray}

The vanishing of the four $\beta ^{i}$ is equivalent to the PDE's
of Eqs.(\ref{pde1}), which motivates their definitions. For later
use, we choose the equivalent set of one-forms,

\begin{eqnarray}
\theta ^{0}\, &=&\Phi \beta ^{0},  \label{theta} \\
\theta ^{+} &=&\Phi \alpha (\beta ^{+}+b\beta ^{-}),  \nonumber \\
\theta ^{-} &=& \Phi \alpha (\beta ^{-}+ b^{*}\beta ^{+}),
\nonumber \\
\theta ^{1}\, &=&\Phi (\beta ^{1}+a\beta ^{+}+a^{*}\beta ^{-}+
c\beta ^{0}). \nonumber
\end{eqnarray}
We refer to the set $(\alpha ,b,b^{*},a,a^{*},c)$ as\emph{\ tetrad
parameters }and $\Phi $ as a conformal parameter. For the moment
they are undetermined functions of $(S,S^{*})$ and their
derivatives. Later, we uniquely determine $(\alpha
,b,b^{*},a,a^{*},c)$ explicitly in terms of $(S,S^{*})$ and impose
conditions on $\Phi $.

\begin{remark}
Note that one could generalize the $\theta ^{i}$ by including more
parameters, i.e., by taking linear combinations of the
$\theta^{i}$. We will not do so, as the above definitions of the
$\theta ^{i}$ are sufficient for our purposes. For the study of
Cartan's equivalence problem for pairs of 2$^{nd}$ order PDE's,
the other parameters are needed. In either case, however,
$\beta^{0}$ must be preserved up to scale. We will return to the
issue of the other parameters in Section V in relationship to the
Cartan normal conformal connection\cite{Ch}.
\end{remark}

\begin{remark}
We impose two different conditions on $\Phi $: (1) for
intermediate or transitional use in the display of complicated
expressions in the appendices we use $\Phi =1$ and refer, in this
case, to the $\theta ^{\prime }$s as $\widehat{\theta }\,^{\prime
}$s; (2) a more basic choice is for $\Phi $ to satisfy a certain
differential equation that simplifies the structure of the
conformal metric defined below. (For this choice see section III.)
We thus have that
\begin{equation}
\theta ^{i}=\Phi \widehat{\theta }^{\,i},  \label{thetaTHETA}
\end{equation}
for the non-trivial $\Phi $.
\end{remark}

From Eq.(\ref{theta}), the dual basis vectors, $e_{i}$, are

\begin{eqnarray}
e_{0}\, &=&\Phi ^{-1}(\partial _{Z}-c\partial _{R}), \label{e's} \\
e_{+}\, &=&\Phi ^{-1}\frac{\partial _{W}-b^{*}\partial
_{W^{*}}-(a-a^{*}b^{*})\partial _{R}}{\alpha (1-bb^{*})}
\nonumber \\
e_{-}\, &=&\Phi ^{-1}\frac{\partial _{W^{*}}-b\partial
_{W}-(a^{*}-ab)
\partial _{R}}{\alpha (1-bb^{*})}  \nonumber \\
e_{1}\, &=&\Phi ^{-1}\partial _{R}.  \nonumber
\end{eqnarray}
From Eq.(\ref{thetaTHETA}),
\begin{equation}
e_{i}=\Phi ^{-1}\widehat{e}_{i}.  \label{ehat2}
\end{equation}

By adding the forms

\begin{eqnarray}
\theta ^{s} &\equiv &ds,  \label{ds} \\
\theta ^{s^{*}} &\equiv &ds^{*},  \nonumber
\end{eqnarray}
which are the duals to the vectors $e$'s, Eq.(\ref{es}), to the
four $\theta ^{i}$ defined above, we have a basis of one-forms on
the six dimensional space $(Z,W,W^{*},R,s,s^{*})$. We will refer
to $\theta^{0}$, $\theta ^{+}$ , $\theta ^{-}$, and $\theta ^{1}$
as the space-time set of one-forms, and we will refer to $\theta
^{s}$ and $\theta ^{s^{*}}$ as fiber one-forms. We denote the
space-time one-forms with the lower-case $i$, $j$, etc., and
denote all six one-forms with the upper-case indices $I$, $J$,
etc. Thus,

\begin{eqnarray}
\theta ^{i} &\in &\{\theta ^{0},\theta ^{+},\theta
^{-},\theta^{1}\},
\label{notatin3} \\
\theta ^{I} &\in &\{\theta ^{0},\theta ^{+},\theta ^{-},\theta
^{1}, \theta ^{s},\theta ^{s^{*}}\}.  \label{notation4}
\end{eqnarray}

Note that, in general, a $p$-form with tetrad indices will have
components in all six dimensions. For example, the one form
$\Pi_{j}^{i}$ and the two-form $\Upsilon ^{i}$ will have the
respective expansions

\begin{eqnarray}
\Pi _{\;j}^{i} &=&\Pi _{\;jK}^{i}\theta ^{K}= \Pi
_{\;jk}^{i}\theta ^{k}+\Pi _{\;js}^{i}\theta ^{s}+\Pi
_{\;js^{*}}^{i}\theta ^{s^{*}}, \label{notation5}
\\
\Upsilon ^{i} &=&\frac{1}{2}\Upsilon _{\;JK}^{i}\theta^{J}\wedge
\theta^{K},  \nonumber \\
&=&\frac{1}{2}\Upsilon _{\;jk}^{i}\theta ^{j}\wedge \theta ^{k}+
\Upsilon _{\;js}^{i}\theta ^{j}\wedge \theta ^{s}+\Upsilon
_{\;js^{*}}^{i}\theta ^{j}\wedge \theta ^{s^{*}}+\Upsilon
_{\;ss^{*}}^{i}\theta ^{s}\wedge \theta ^{s^{*}},  \nonumber
\end{eqnarray}

For later use, we construct a metric to make $\theta^{i}$ a null
tetrad

\begin{eqnarray}
g(Z,W,W^{*},R,s,s^{*}) &=& \theta ^{0}\otimes \theta ^{1}+\theta
^{1}\otimes \theta ^{0}-\theta ^{+}\otimes \theta ^{-}-\theta
^{-}\otimes \theta ^{+},
\label{g} \\
&=&\eta _{ij}\theta ^{i}\otimes \theta ^{j}.  \nonumber
\end{eqnarray}
This defines the constant flat metric $\eta _{ij}$ as

\begin{equation}
\eta _{ij}=\left[
\begin{array}{rrrr}
0 & 1 & 0 & 0 \\
1 & 0 & 0 & 0 \\
0 & 0 & 0 & -1 \\
0 & 0 & -1 & 0
\end{array}
\right] .  \label{flatmet}
\end{equation}
From Eq.(\ref{inequal}), it follows\cite{KN,FKN} that the metric
$g$ is Lorentzian.

In addition, we also define the symmetric tensors $G_{ij}$ and
$G_{ij}^{*}$ from the Lie derivatives of the metric in the $e_{s}$
and $e_{s^{*}}$ directions:
\begin{eqnarray}
\pounds _{e_{s}}g &=&G_{ij}\theta ^{i}\otimes \theta ^{j},
\label{Lieg} \\
\pounds _{e_{s^{*}}}g &=&G_{ij}^{*}\theta ^{i}\otimes \theta ^{j},
\nonumber
\end{eqnarray}
and the exterior derivatives of the one-form basis by

\begin{equation}
d\theta ^{i}= \frac{1}{2}\triangle ^{i}\,_{JK}\theta ^{J}\wedge
\theta ^{K}. \label{triangle*}
\end{equation}
From this and
\begin{equation}
\pounds _{e_{s}}\theta ^{i}=e_{s}\rfloor \,d\theta ^{i},
\end{equation}
we have that the $G_{ij}$ and the $\triangle^{i}\,_{JK}$ are
related by

\begin{eqnarray}
G_{ij} &=&-2\triangle \,_{(ij)s},  \label{G,triangle} \\
G_{ij}^{*} &=&-2\triangle \,_{(ij)s^{*}},  \nonumber
\end{eqnarray}
where
\begin{equation}
\triangle \,_{iJK}=\eta _{il}\triangle ^{l}\,_{JK}.
\end{equation}
The expressions for both $G_{ij}$ and $\triangle^{i}\,_{JK},$ in
terms of $ S,S^{*},\Phi ,$ the tetrad parameters and their
derivatives, are given in Appendix B.

\newpage

\section{The First Structure Equation}

We begin by inserting our one-forms, $\theta^{i}\in \{\theta
^{0},\theta ^{+},\theta ^{-},\theta ^{1}\}$, into Cartan's
torsion-free first structure equation,

\begin{equation}
d\theta ^{i}+\omega _{\;j}^{i}\wedge \theta ^{j}=0.
\label{structureone}
\end{equation}
Our goal now is to solve this equation for the connection
one-forms, $\omega^{i}\,_{j}$. To do so, write

\begin{eqnarray}
d\theta ^{i} &=&\frac{1}{2}\Delta _{\;JK}^{i}\theta^{J}\wedge
\theta ^{K},
\label{triangle} \\
\omega _{ij} &=&\omega _{ijK}\theta ^{K},  \label{spincoef}
\end{eqnarray}
defining the $\triangle ^{i}\,_{JK}$ and $\omega_{ijK}$. Note that
\begin{equation}
\omega ^{i}\,_{k}=\eta ^{ij}\omega _{jk},  \label{def}
\end{equation}
where $\eta_{ij}$ is the flat metric defined in
Eq.(\ref{flatmet}). The structure equation then becomes
\begin{equation}
\frac{1}{2}\triangle ^{i}\,_{JK}\theta ^{J}\wedge \theta ^{K}+
\eta ^{ij}\omega _{jkL}\theta ^{L}\wedge \theta ^{k}=0.
\label{structuretwo}
\end{equation}

Since we are interested the conformal geometry contained in the
structure equation, we require that the connection one-forms be
generalized Weyl connections (``generalized'' because of the extra
degrees of freedom in the fiber directions, $s$ and $s^{*}$):
\begin{eqnarray}
\omega _{ij} &=&\omega_{[ij]}+\omega_{(ij)},
\label{weylconnone} \\
\omega _{(ij)} &=&\eta _{ij}A,  \nonumber
\end{eqnarray}
where the one-form
\begin{equation}
A=A_{I}\theta ^{I}=A_{i}\theta ^{i}+A_{s}\theta ^{s}+
A_{s^{*}}\theta ^{s^{*}},  \label{defA}
\end{equation}
is the (generalized) Weyl one-form.

In Eqs.(\ref{theta}) we expressed our space-time tetrad, $\theta
^{i}$, in terms of $S$, $S^{*}$, and the unspecified tetrad
parameters, $(\alpha ,b,b^{*},a,a^{*},c)$ and $\Phi $. Thus, we
can explicitly compute the $ \triangle ^{i}\,_{JK}$ in terms of
$S$, $S^{*}$, the tetrad parameters, $\Phi $ and their
derivatives. (The explicit expressions for the $\triangle
_{JK}^{i}$, as we said earlier, are given in Appendix B.)
Therefore, we will use Eq.(\ref{structuretwo}) to solve for the
connection coefficients, $\omega _{ijK}$, in terms of the
$\triangle ^{i}\,_{JK}$ and the undetermined $A_{I}$. In doing so,
we will find several things: \emph{i)} the four space-time
components of the Weyl-one form, $A_{i}$, remain arbitrary;
\emph{\ \ \ ii)} the skew-symmetric part of the connection,
$\omega _{[ij]}$, and the fiber parts of the Weyl-one form,
$A_{s}$ and $A_{s^{*}}$, are uniquely determined functions of $S$,
$S^{*}$, $A_{i}$\thinspace and $\Phi $; \emph{\ iii)} the tetrad
parameters are uniquely determined functions of $S$ and $S^{*}$;
and \emph{iv)} there must be restrictions on the class of 2$^{nd}$
order PDE's to which $S$ and $S^{*}$ belong. The conditions of
\emph{(iv)} are known as the (generalized) W\"{u}nschmann
condition and its complex-conjugate. (Note that when we say
``W\"{u}nschmann condition'', we often mean both the condition and
its conjugate.) These conditions are complex differential
equations in all six variables of our six-dimensional space, ($Z$,
$W$, $W^{*}$, $R$, $s$ , $s^{*}$).

We begin by splitting the structure equation it into its
fiber-fiber, tetrad-fiber, and tetrad-tetrad components.

A. The fiber-fiber component contains no information since the
term $\omega_{ijK}\theta^{K}\wedge \theta ^{j}$ has no fiber-fiber
parts and, (using direct calculation and integrability conditions)

\begin{equation}
\triangle ^{i}\,_{ss^{*}}\equiv 0.  \label{trifiberfiber}
\end{equation}

B. The tetrad-fiber parts of the structure equation are

\begin{eqnarray}
\omega _{ijs} &=&\triangle \,_{ijs},  \label{omegaijs} \\
\omega _{ijs^{*}} &=&\triangle \,_{ijs^{*}}.  \nonumber
\end{eqnarray}

An important observation to make is that

\begin{equation}
\triangle \,_{ijs}=\widehat{\triangle }\,_{ijs}-\eta
_{ij}\Phi^{-1}D\Phi , \label{triangle.hat}
\end{equation}
where $\widehat{\triangle }\,_{js}^{k}$ is defined by
\begin{equation}
d\widehat{\theta }^{i}=\frac{1}{2}\widehat{\Delta
}_{\;JK}^{i}\widehat{ \theta }^{J}\wedge \widehat{\theta }^{K}.
\label{dthetahat}
\end{equation}

Symmetrizing on ($i,j)$ in Eq.(\ref{omegaijs}) and using
Eqs.(\ref{weylconnone}) and (\ref{triangle.hat}) yields
\begin{eqnarray}
\eta _{ij}A_{s} &=&\triangle \,_{(ij)s}=\widehat{\triangle
}\,_{(ij)s}- \eta
_{ij}\Phi ^{-1}D\Phi ,  \label{As} \\
\eta _{ij}A_{s^{*}} &=&\triangle \,_{(ij)}=\widehat{\triangle }
\,_{(ij)s^{*}}-\eta _{ij}\Phi ^{-1}D^{*}\Phi ,  \nonumber
\end{eqnarray}
while the skew-symmetric parts gives

\begin{eqnarray}
\omega _{[ij]s} &=&\triangle \,_{[ij]s}= \widehat{\triangle
}\,_{[ij]s},
\label{ws2} \\
\omega _{[ij]s^{*}} &=&\triangle \,_{[ij]s^{*}}=
\widehat{\triangle } \,_{[ij]s^{*}}.  \nonumber
\end{eqnarray}

Eqs.(\ref{As}), uniquely determine $A_{s}$ and $A_{s^{*}}$ in
terms of $S$, $S^{*}$, and $\Phi$ as
\begin{eqnarray}
A_{s} &=&\frac{1}{4}\triangle ^{k}\,_{ks}=\widehat{A}_{s}-\Phi
^{-1}D\Phi,
\label{A-s} \\
A_{s*} &=&\frac{1}{4}\triangle ^{k}\,_{ks*}=\widehat{A}_{s*} -
\Phi
^{-1}D^{*}\Phi ,  \nonumber \\
\widehat{A}_{s} &\equiv &\frac{1}{4} \widehat{\triangle
}^{k}\,_{ks}. \label{A-shat}
\end{eqnarray}

In addition, the trace-free part of Eqs.(\ref{As}),
\begin{eqnarray}
\triangle \,_{(ij)s}-\frac{1}{4}\eta _{ij}\triangle ^{k}\,_{ks}
&=& \widehat{ \triangle }\,_{(ij)s}-\frac{1}{4}\eta
_{ij}\widehat{\triangle }^{k}\,_{ks}=0,
\label{TrFree} \\
\triangle \,_{(ij)s^{*}}-\frac{1}{4}\eta _{ij}\triangle
^{k}\,_{ks^{*}} &=& \widehat{\triangle
}\,_{(ij)s^{*}}-\frac{1}{4}\eta _{ij}\widehat{\triangle }
^{k}\,_{ks^{*}}=0,  \nonumber
\end{eqnarray}
imposes conditions on the $S$, $S^{*}$ and determines uniquely the
tetrad parameters while $\Phi $ remains undetermined.
Alternatively, from Eq.(\ref{G,triangle}), i.e.,
$G_{ij}=-2\triangle _{(ij)s},$ and by the definition
\begin{equation}
\widehat{G}_{ij}=-2\widehat{\triangle }_{(ij)s},  \label{Ghat}
\end{equation}
we have
\begin{equation}
G_{ij} = \widehat{G}_{ij}+2\eta _{ij}\Phi ^{-1}D\Phi ,
\label{GGhat}
\end{equation}
from which we easily see that
\begin{equation}
G_{ij}^{TF}=\widehat{G}_{ij}^{TF}=0,  \label{GhatTF}
\end{equation}
where TF denotes the trace-free part.

The details for analyzing Eqs.(\ref{GhatTF}) are quite involved
and will be given in appendix C.

\begin{theorem}
From Eqs.(\ref{As}) and the relationship between the
$\bigtriangleup _{\,js}^{k}$and{\ }the $G_{ij}$, i.e.,
Eqs.(\ref{G,triangle}), we have the result
\begin{eqnarray}
\pounds _{e_{s}}g &=&-2A_{s}g,  \label{th1} \\
\pounds _{e_{s^{*}}}g &=&-2A_{s^{*}}g.  \nonumber
\end{eqnarray}
It follows from this theorem that the degenerate six-dimensional
metric defined in Eq.(\ref{g}) yields a conformal four-metric in
the solution space, with{\ the motion along either $e_{s}$ or
$e_{s^{*}}$ generating the conformal re-scaling of the metric.}
\end{theorem}

\begin{corollary}
Up to this point $\Phi $ has remained undetermined. There is
however, via Eq.(\ref{A-s}), a canonical way to chose it, namely
by,
\begin{eqnarray}
D\Phi -\frac{1}{4}\widehat{\triangle }^{k}\,_{ks}\Phi  &=&0,
\label{Dphi} \\
D^{*}\Phi -\frac{1}{4}\widehat{\triangle }^{k}\,_{ks^{*}}\Phi
&=&0, \nonumber
\end{eqnarray}
making $A_{s}=A_{s^{*}}=0.$ We thus have that{\
\begin{eqnarray}
\pounds _{e_{s}}g &=&0,  \label{Lg=0} \\
\pounds _{e_{s^{*}}}g &=&0.  \nonumber
\end{eqnarray}
}
\end{corollary}

\begin{remark}
Note that the freedom in the solution of Eq.(\ref{Dphi}) for $\Phi
$ is an arbitrary multiplicative function (referred to as the
conformal factor) $ \varpi $ such that $D\varpi =D^{*}\varpi =0,$
i.e., $\varpi $ is an arbitrary function on the space of fibers
(the solution space, $\frak{M}^{4}$ ). The solution is thus of the
form $\Phi =\varpi \Phi _{0}$ where
\[
\Phi _{0}=\Phi _{0}[S,S^{*}]=\exp \frac{1}{4}(\int
\widehat{\triangle} ^{k}\,_{ks}ds+\widehat{\triangle
}^{k}\,_{ks^{*}}ds^{*}).
\]
where the integral is taken along an arbitrary path from any
initial point to the final point ($s,s^{*}$). The needed
integrability conditions are satisfied. Multiplication of the
metric by $\varpi ^{2}$ is the ordinary conformal freedom,
$g\Rightarrow \varpi ^{2}g.$
\end{remark}

C. Returning to the tetrad-tetrad parts of the structure
equations, we have that
\begin{equation}
\triangle ^{i}\,_{mn}+\eta ^{ij}(\omega _{jnm}-\omega _{jmn})=0,
\label{tt}
\end{equation}
or
\begin{equation}
\omega _{i[jk]}=\frac{1}{2}\triangle \,_{ijk}. \label{omegaijkone}
\end{equation}

From the tensor identity
\begin{equation}
\omega _{ijk} = \omega _{(ij)k}-\omega _{(jk)i}+\omega
_{(ki)j}+\omega _{i[jk]}-\omega _{k[ij]}+\omega _{j[ki]},
\label{identity}
\end{equation}
and Eqs.(\ref{weylconnone}) and (\ref{omegaijkone}), we obtain the
tetrad-tetrad coefficients of the connection;

\begin{equation}
\omega _{ijk} = \eta _{ij}A_{k}-\eta _{jk}A_{i}+\eta
_{ki}A_{j}+\frac{1}{2} (\eta _{mi}\triangle ^{m}\,_{jk}-\eta
_{mk}\triangle ^{m}\,_{ij}+\eta _{mj}\triangle ^{m}\,_{ki}).
\label{wijk}
\end{equation}
This decomposes naturally into a Levi-Civita part $\gamma_{ijk}$
(which is independent of $A_{i}$) plus a ``Weyl'' part,
$\tilde{\omega}_{ijk},$ i.e.,
\begin{eqnarray}
\omega _{ijk} &=& \gamma _{ijk}+\tilde{\omega}_{ijk},
\label{omegijk} \\
\gamma _{ijk} &=& \frac{1}{2}(\eta _{mi}\triangle ^{m}\,_{jk}-\eta
_{mk}\triangle ^{m}\,_{ij}+\eta _{mj}\triangle ^{m}\,_{ki}),
\label{gammaijk} \\
\tilde{\omega}_{ijk} &=& \eta _{ij}A_{k}-\eta _{jk}A_{i}+\eta
_{ki}A_{j}. \label{w;hat}
\end{eqnarray}

For clarity of exposition we point out where different variables
are hidden. First note that $\triangle_{imn}$ can be decomposed,
as in Eq.(\ref {triangle.hat}), as

\begin{equation}
\triangle \,_{imn}=\widehat{\triangle }\,_{imn}\Phi ^{-1}+
2\widehat{e} _{[m}\Phi \cdot \eta _{n]\,\,i}\Phi ^{-2},
\label{2triangle}
\end{equation}
where $\widehat{\triangle }\,_{imn}$ is the same expression as
$\triangle _{imn}$ but with $\Phi =1.$ The $\widehat{\triangle
}\,_{imn}$ depends only on the ($S,S^{*})$. This means that
$\gamma _{ijk},$ Eq.(\ref{gammaijk}), also decomposes into terms
that depend on the ($S,S^{*})$ and the $\Phi$. Since $\Phi =\varpi
\Phi _{0}[S,S^{*}],$ with $\varpi =\varpi (x^{a}),$ an arbitrary
function of $x^{a},$ carried along. This gauge or conformal
freedom will be discussed later.

In summary, we have shown that Eqs.(\ref{omegaijs}) and
(\ref{wijk}) completely determine the $\omega _{ijK}$ in terms of
the $\triangle^{i}\,_{JK}$ and the undetermined space-time parts
of $A$, i.e., $A_{i}$.

\subsection{A Theorem}

To conclude this section we return to the vanishing of the
trace-free part of $\triangle \,_{(ij)s}$ or the trace-free part
of $\widehat{\triangle}\,_{(ij)s}$, i.e., to Eq.(\ref{TrFree}).
They are nine complex equation for the determination of the tetrad
parameters, ($\alpha ,a,a^{*},b,b^{*},c$). Thus either there must
be several identities and/or conditions to be imposed on the $S$
and $S^{*}.$ By explicitly solving these equations, (see Appendix
C), the results can be summarized in the following theorem:

\begin{theorem}
The torsion free condition on the connection:

1. Uniquely determines the connection $\omega_{ij},$ via
Eqs.(\ref{wijk} ) and (\ref{omegaijs}).

2. Uniquely determines the tetrad parameters in terms of $S$ and
$S^{*},$ (see below).

3. Imposes a (complex) condition, the vanishing of the
W\"{u}nschmann invariant, on $S$ and $S^{*},$ (see below) with the
tetrad parameters given by
\begin{eqnarray}
b &=&\frac{-1+\sqrt{1-S_{R}S_{R}^{*}}}{S_{R}^{*}},  \label{b} \\
\alpha ^{2} &=&\frac{1+bb^{*}}{(1-bb^{*})^{2}},  \label{alpha}
\end{eqnarray}
\begin{eqnarray}
a &=&b^{-1}b^{*-1}(1-bb^{*})^{-2}(1+bb^{*})\{b^{*2}(-Db+bS_{W}-
S_{W^{*}})
\label{a} \\
&&+b(-D^{*}b^{*}+b^{*}S_{W^{*}}^{*}-S_{W}^{*})\},  \nonumber
\end{eqnarray}
\begin{eqnarray}
c &=& -\frac{Da + D^{*}a^{*}+T_{W}+T_{W^{*}}^{*}}{4} -
\frac{aa^{*}(1+6bb^{*}+b^{2}b^{*2})}
{2(1+bb^{*})^{2}}  \label{c} \\
&& +\frac{(1+bb^{*})(bS_{Z}^{*}+b^{*}S_{Z})}{2(1-bb^{*})^{2}} +
\frac{a(2ab-b^{*}S_{W^{*}})+
a^{*}(2a^{*}b^{*}-bS_{W}^{*})}{2(1+bb^{*})}, \nonumber
\end{eqnarray}
and the differential (W\"{u}nschmann) condition imposed on $S$ and
$S^{*}$ is
\begin{equation}
M\equiv \frac{Db + bD^{*}b+ S_{W^{*}}-bS_{W}+
b^{2}S_{W^{*}}^{*}-b^{3}S_{W}^{*}}{1-bb^{*}}= 0.
\label{Wunschmann}
\end{equation}
\end{theorem}

\section{The Cartan Curvatures}

In the previous section, we used the first structure equation,
Eq.(\ref{structureone}), to algebraically solve for the components
of a torsion-free connection defined with the symmetry
\begin{equation}
\omega _{ij}=\omega _{[ij]}+\eta _{ij}A,  \label{con}
\end{equation}
uniquely in terms of $(S,S^{*})$ and the undetermined $A_{i}$ and
$\varpi$.

Our next goal is to compute the curvature two-forms,
$\Theta_{ij}$, defined by the \emph{second structure equation}:
\begin{equation}
d\omega ^{i}\,_{j}+\omega ^{i}\,_{k}\wedge \omega ^{k}\,_{j}=
\Theta ^{i}\,_{j}=\frac{1}{2}\Theta ^{i}\,_{jLM}\theta ^{L}\wedge
\theta ^{M}. \label{secondstruct}
\end{equation}

By taking the exterior derivative of the first structure equation,
Eq.(\ref {structureone}), with the second structure equation,
Eq.(\ref{secondstruct}), we obtain the \emph{first Bianchi
identity}:
\begin{equation}
\Theta _{ij}\wedge \theta ^{j}=0\quad \Leftrightarrow \quad \Theta
_{i[jLM]}=0.  \label{BI}
\end{equation}

Splitting this into its tetrad-tetrad, tetrad-fiber, and
fiber-fiber parts, we have

\begin{eqnarray}
\Theta _{ijkm}+\Theta _{ikmj}+\Theta _{imjk} &=&0,
\label{firtettetbianchi}
\\
\Theta _{i[jk]s} &=&0,  \label{asdf} \\
\Theta _{ijss^{*}} &=&0,  \label{curvfibfib}
\end{eqnarray}
The last two relations are due to the fact that the first two
indices of $\Theta _{ijLM}$ are only $4$-dimensional, whereas the
last two are $6$-dimensional.

We calculate the $\Theta _{ij}$ as explicit functions of
$(S,S^{*})$ and the undetermined $A_{i}$ and $\varpi $. First,
note that from Eqs.(\ref{con}) and (\ref{secondstruct}), it is
straightforward to see that the $\Theta_{ij} $ inherits the
symmetry of the $\omega _{ij}$, and thus can be written as
\begin{equation}
\Theta _{ij}=\Theta _{[ij]}+\eta _{ij}dA,  \label{curvsymmetry}
\end{equation}
with
\begin{equation}
dA=\frac{1}{2}(dA)_{LM}\theta ^{L}\wedge \theta ^{M}.
\end{equation}
(This defines the components $(dA)_{LM}$.)

Next, we split the components $\Theta _{ijLM}$ into tetrad-tetrad
parts, $ \Theta _{ijkm}$, and tetrad-fiber parts, $\Theta
_{ijks}$. (The fiber-fiber parts are identically zero from the
first Bianchi identity). We first calculate the $\Theta _{ijkm}$,
which can be split into terms arising from the Levi-Civita part of
the connection and terms arising from the Weyl part of the
connection. These are denoted respectively by $\Re _{[ij][km]}$
and $\tilde{\Theta}_{ij[km]}$,

\begin{eqnarray}
\Theta _{ij[km]} &=& \Re _{[ij][km]}+\tilde{\Theta}_{ij[km]},
\label{CurveT}
\\
&=&\Re _{[ij][km]}+\tilde{\Theta}_{[ij][km]}+\eta _{ij}(dA)_{[km]}.
\nonumber
\end{eqnarray}
The $\Re _{[ij][km]}$ are the components of the standard Riemann
tensor of the (Levi-Civita) $\gamma_{ijk}$ connection.

The $\tilde{\Theta}_{[ij][km]}$ depends on $A$ and its
derivatives. Denoting the covariant derivative associated with the
Levi-Civita connection $\gamma_{ijk},$ by $\nabla _{i},$ we have
\begin{equation}
\nabla _{i}A_{j}=e_{i}(A_{j})-\gamma _{kji}A^{k},  \label{consplit}
\end{equation}
and
\begin{equation}
(dA)_{ij}=2\nabla _{[i}A_{j]}.  \label{chi}
\end{equation}
$\tilde{\Theta}_{[ij][km]}$ can, thus, be written as

\begin{eqnarray}
\frac{1}{2}\tilde{\Theta}_{[ij][km]} &=& \eta_{j[k}
\nabla_{m]}A_{i} - \eta_{i[k} \nabla_{m]}A_{j} + A^{2}
\eta_{j[k} \eta_{m]\,\,i}  \label{THETAhat} \\
&&+A_{j}\eta _{i[k}A_{m]}-A_{i}\eta _{j[k}A_{m]},  \nonumber
\end{eqnarray}
where $A^{2}=A^{m}A_{m}.$

Defining
\begin{equation}
R_{jm}\equiv \eta ^{ik}\Theta _{ijkm},  \label{defrictensor}
\end{equation}
and using Eqs.(\ref{CurveT}) and (\ref{THETAhat}), we obtain
\begin{equation}
R_{jm}=\Re _{(jm)}-\eta _{jm}\nabla _{p}A^{p}-2\{\nabla
_{(m}A_{j)}+ \eta _{jm}A^{2}-A_{j}A_{m}\}+4\nabla _{[j}A_{m]},
\label{Rij}
\end{equation}
where the $\frak{R}_{(jm)}$ are the components of the Ricci tensor
of the $\gamma _{ijk}$.

If we also let
\begin{equation}
R\equiv \eta ^{jm}R_{jm},  \label{defricscalar}
\end{equation}
then from Eq.(\ref{Rij}), we obtain
\begin{equation}
R=\Re -6\{\nabla _{p}A^{p}+A^{2}\},  \label{ricciscalar}
\end{equation}
where $\frak{R}$ is the standard Ricci scalar.

The tetrad-fiber part of $\Theta _{ij}$, which is

\begin{equation}
\Theta _{ijks} = \eta _{ij}(dA)_{ks}+\eta _{ik}(dA)_{js}-\eta
_{jk}(dA)_{is}, \label{THETAS}
\end{equation}
will be derived below.

\subsection{The First Cartan Curvature}

The Cartan first curvature two-form is given by
\begin{equation}
\Omega _{ij}= \Theta _{ij}+\Psi _{i}\wedge \eta _{jk}\theta
^{k}+\eta _{ik}\theta ^{k}\wedge \Psi _{j}-\eta _{ij}\Psi
_{k}\wedge \theta ^{k}, \label{carwithpsi}
\end{equation}
where the (Ricci) one-forms $\Psi_{i}$ are appropriately chosen so
that
\begin{equation}
\Omega _{ij} = \frac{1}{2}\Omega _{ijLM}\theta ^{L}\wedge \theta
^{M}, \label{OMEGA}
\end{equation}
satisfies the following conditions

\begin{eqnarray}
\Omega _{ijkm} &=&\Omega _{[ij]km},  \label{fircarskew} \\
\eta ^{ik}\Omega _{ijkm} &=&0,  \label{traceonethree} \\
\Omega _{ijks} &=&0.  \label{fircartetfib}
\end{eqnarray}
Note, from its definition, that $\Omega_{ij}$ also satisfies the
first Bianchi identity, Eq.(\ref{BI}), i.e.,
\begin{equation}
\Omega _{ij}\wedge \theta ^{j}=0.  \label{fircarbian}
\end{equation}

It is straightforward to show that the conditions
Eqs.(\ref{fircarskew}),(\ref{traceonethree}) and
(\ref{fircartetfib}) are satisfied uniquely by the one-form
\begin{equation}
\Psi _{i}=\Psi _{iK}\theta ^{K}= \Psi _{ij}\theta ^{j}+\Psi
_{is}\theta ^{s}+\Psi _{is^{*}}\theta ^{s^{*}},  \label{PSIi}
\end{equation}
with
\begin{equation}
\Psi _{ij} = -\frac{1}{4}R_{[ij]} - \frac{1}{2}(R_{(ij)} -
\frac{1}{6}R\eta _{ij}). \label{psi}
\end{equation}
and

\begin{eqnarray}
\Psi _{is} &=&-(dA)_{is},  \label{PSIs} \\
\Psi _{is^{*}} &=&-(dA)_{is^{*}}.  \nonumber
\end{eqnarray}

From Eqs.(\ref{PSIs}), (\ref{carwithpsi}) and
(\ref{fircartetfib}), we find Eq.(\ref{THETAS}),i.e.,

\[
\Theta _{ijks}= \eta _{ij}(dA)_{ks}+\eta _{ik}(dA)_{js}-\eta
_{jk}(dA)_{is}.
\]

Using Eqs.(\ref{Rij}) and (\ref{ricciscalar}), we obtain

\begin{equation}
\Psi _{ij}=\Im _{ij}-\nabla _{[i}A_{j]}-2\{\nabla
_{(i}A_{j)}+\frac{1}{2} \eta _{ij}A^{2}-A_{i}A_{i}\},
\label{psi2}
\end{equation}
with
\begin{equation}
\Im _{ij}=-\frac{1}{2}(\Re _{(ij)}-\frac{1}{6}\Re \eta _{ij}).
\label{Sij}
\end{equation}

By using Eq.(\ref{psi2}), we can insert the above expression into
Eq. (\ref{carwithpsi}), yielding
\begin{equation}
\Omega _{ijkm}= \Re _{ijkm}-\eta _{kj}\Im _{im}+\eta _{ki}\Im
_{ijm}-\eta _{mi}\Im _{jk}+\eta _{mj}\Im _{ik},  \label{curvijkm}
\end{equation}
which is manifestly independent of the $A_{i}$. Furthermore, using
Eq.(\ref{Sij}), we recover the standard definition of the Weyl
tensor,
\begin{equation}
\Omega _{ijkm}=C_{ijkm}.  \label{weyl}
\end{equation}

\subsection{Second Cartan Curvature}

Finally, we define the second Cartan curvature (with the covariant
exterior derivative $\mathfrak{D}$) of the one-form $\Psi_{i}$ as
\begin{eqnarray}
\Omega _{i} &=&d\Psi _{i}+\Psi _{k}\wedge \omega^{k}\,_{i} \equiv
\mathfrak{D}\Psi _{i},  \label{thirdstruct} \\
&=&\frac{1}{2}\Omega _{iJK}\theta ^{J}\wedge \theta ^{K}.
\nonumber
\end{eqnarray}
Using Eq.(\ref{PSIi}) in the above, we obtain, after a lengthy
calculation, the simple results

\begin{equation}
\Omega _{imn}=\nabla ^{j}C_{ijmn}+A^{j}C_{ijmn}, \label{OMEGAijk}
\end{equation}
and
\begin{eqnarray}
\Omega _{ims} &=&0,  \label{OMEGAis} \\
\Omega _{iss^{*}} &=&0.  \nonumber
\end{eqnarray}
$\nabla ^{j}$ again is the Levi-Civita covariant derivative.

In section VI, the two Cartan curvatures will be used to construct
the curvature of a conformal normal Cartan connection.

\section{Synopsis}

Since so many different quantities and their symbols have been
introduced, we have added a few essentially pedagogical remarks
concerning the placement of different variables and where the
conformal transformation acts.

First we return to the definitions of $\theta^{i},$ i.e.,
Eq.(\ref{theta}) and write them (and their related duals) as

\begin{eqnarray}
\theta ^{i} &=& \Phi \,\widehat{\theta }^{i},
\label{thetathetahat} \\
e_{i} &=&\Phi ^{-1}\,\widehat{e_{i}},  \label{eehat}
\end{eqnarray}
and observe that $\bigtriangleup_{\,\,JK}^{i}$ and
$\widehat{\bigtriangleup }_{\,\,JK}^{i}$ and their relationship,
Eqs.(\ref{triangle.hat}) and (\ref {2triangle}), arise from
\begin{equation}
d\,\theta ^{i}=\frac{1}{2}\bigtriangleup _{\,\,JK}^{i}\theta
^{j}\wedge \theta ^{K}\text{ }\,\text{\& }\,\text{
}d\,\widehat{\theta }^{i}=\frac{1}{2} \widehat{\bigtriangleup
}_{\,\,JK}^{i}\,\widehat{\theta }^{j}\wedge \widehat{ \theta
}^{K}.  \label{dtheta}
\end{equation}
The action of $\Phi,$ taking $\widehat{\theta }^{i}\Rightarrow
\theta ^{i},$ takes the metric
\begin{equation}
\widehat{g}=\eta _{ij}\widehat{\theta }^{i}\otimes \widehat{\theta
} ^{j}=>g=\Phi ^{2}\widehat{g},  \label{ghat}
\end{equation}
where $g$ and $\widehat{g}$ both depend, in general, on
$(s,s^{*}).$ However when we make the special choice $\Phi=\varpi
\Phi _{0},$ Eq.(\ref{Dphi}), the resulting $g$ is then a function
on $\frak{M}^{4}$ alone. What remains is the standard conformal
freedom that is given by the choice of $\varpi (x^{a}).$ Whenever,
in any expression, $\varpi (x^{a}),$ is changed,
\begin{equation}
\varpi (x^{a})\Rightarrow f(x^{a})\,\varpi (x^{a}),
\label{conformalTrans}
\end{equation}
that change constitutes the effect of the conformal transformation.

All the further geometric quantities (the connection and different
curvatures) developed and defined via the structure equations,
contain the following quantities: our basic variables ($S,S^{*})$,
the four arbitrary space-time components of the Weyl one-form
$A=A_{i}\theta ^{i}$ where $A_{s}=0$, and the conformal factor
$\varpi (x^{a}).$

We give a brief survey of where these quantities appear:

a. All $\widehat{\bigtriangleup }_{ijK}$ and
$\bigtriangleup_{ijs}$ depend only on ($S,S^{*}),$ while
$\bigtriangleup _{ijk}$depends on ($ S,S^{*},\varpi ).$ It is an
easy task to see how $\bigtriangleup _{ijk}$ transforms when
$\varpi $ is changed.

b. Since $\omega _{ijs}=\triangle _{ijs},$ it thus depend only on
($S,S^{*}).$

c. All the quantities ($S,S^{*},A_{i},\varpi)$ appear in the
connection one-forms $\omega _{ijk},$ which can be split into

\begin{equation}
\omega_{ijk}=\gamma _{ijk}+\tilde{\omega}_{ijk}, \label{gamma+w}
\end{equation}
where the Levi-Civita part, $\gamma _{ijk},$ depends only on
($S,S^{*},\varpi )$ while $\tilde{\omega}_{ijk}$ depends only on
the $A_{i}.$

d. The curvature $\Theta _{ij[km]}$ splits into two parts

\begin{equation}
\Theta _{ij[km]}=\Re _{[ij][km]}+\tilde{\Theta}_{ij[km]},
\label{whereR}
\end{equation}
where the (standard) Riemann curvature $\Re _{[ij][km]}$ depends
on ($S,S^{*},\varpi )$ and $\tilde{\Theta}_{ij[km]}$ depends on
everything.

e. The first Cartan curvature two-form, $\Omega_{ij}$, \textit{is}
the Weyl tensor, $C_{ijmn},$ and depends only on
($S,S^{*},\varpi).$

f. The second Cartan curvature two-form,

\begin{equation}
\Omega _{i}= \frac{1}{2}\{\nabla
^{m}C_{imjk}+A^{m}C_{imjk}\}\theta ^{j}\wedge \theta ^{k},
\label{secondCC}
\end{equation}
depends on everything, though the $A_{i}$ appears explicitly  just
in the linear term.

g. Though the Ricci one-forms,
\begin{equation}
\Psi _{i}=\Psi _{ij}\theta ^{j}+\Psi _{is}\theta
^{s}+\Psi_{is^{*}}\theta ^{s^{*}},  \label{Riccione}
\end{equation}
depend on everything, their separate parts do not. $\Psi_{is}$
depends only on the $A_{i}.$ From
\begin{equation}
\Psi _{ij}=\Im _{ij}-\nabla _{[i}A_{j]}-2\{\nabla
_{(i}A_{j)}+\frac{1}{2} \eta _{ij}A^{2}-A_{i}A_{i}\},
\label{moreRicci}
\end{equation}
we have that $\Im _{ij}$ depends only on ($S,S^{*},\varpi)$ while
the remaining terms depends on everything.

Whenever a geometric quantity depended only on ($S,S^{*},\varpi)$
we could have said it depended on the Levi-Civita connection
obtained from the metric $g=\varpi^{2}\Phi _{0}^{2}\,\widehat{g},$
Eq.(\ref{g}), and referred to it as a conformal covariant.

\newpage

\section{Unification: Cartan's Normal Conformal Connection}

From a pair of 2$^{nd}$ order PDE's satisfying the W\"{u}nschmann
condition we derived a rich geometric structure on the
four-dimensional solution space of the PDE's. This structure
includes: a conformal metric, a torsion-free connection and
several different curvature tensors. Though it is not obvious, and
the starting point of view is quite different, we are following
Kobayashi's\cite{Kob} development of Cartan's theory of normal
conformal connections via the three structure equations,
Eqs.(\ref{structureone}), (\ref{secondstruct}) and
(\ref{thirdstruct}) and the Ricci one-forms, Eq.(\ref {PSIi}).

We now show that, \textit{basically}, we have, in fact, recovered
a (15 dimensional) principle bundle $P$ over $\frak{M}^{4}$ with
group $ H=CO(1,3)\otimes _{s}T^{*}$ and a Cartan normal conformal
connection with values in the Lie algebra of $O(4,2).$ The group
$H=CO(1,3)\otimes _{s}T^{*}$ is an 11-dimensional subgroup of
$O(4,2)$\cite{Kob} with $T^{*}$ a four-dimensional translation
group. More specifically we have recovered a six-dimensional
subbundle, $J^{6},$ of this 15-dimensional Cartan bundle. The base
space is the four-dimensional solution space $\frak{M}^{4}$ and
the two-dimensional fibers are formed by the integral curves of
$D$ and $D^{*}$.

We began with the pair of 2nd order PDE's (that satisfy the
W\"{u}nschmann condition) and a set of four associated one-forms,
$\theta^{i},$ plus the two fiber one-forms, $(ds,ds^{*}),$ on the
six-dimensional space $J^{6}.$ We then found the connection

\[
\omega _{ij}=\omega _{[ij]}+A\eta _{ij},
\]
satisfying the first (torsion-free) structure equation

\begin{equation}
d\theta ^{i}+\omega _{\;j}^{i}\wedge \theta ^{j}=0,  \label{ONE}
\end{equation}
where $A_{s}=A_{s^{*}}=0$ and the four $A_{i}$ are arbitrary. The
first-Cartan curvature two-form was found via the second structure
equation

\begin{eqnarray}
\Omega _{ij} &=& d\omega _{ij}+\eta ^{kl}\omega _{ik}\wedge \omega
_{lj}+\eta _{il}\theta ^{l}\wedge \Psi _{j}+\Psi _{i}\wedge \theta
^{l}\eta _{jl}-\eta
_{ij}\Psi _{k}\wedge \theta ^{k}  \label{TWO} \\
&=&\frac{1}{2}C_{ijlm}\theta ^{l}\wedge \theta ^{m},
\end{eqnarray}
with the proper choice of the Ricci one-forms $\Psi_{i}$,
Eq.(\ref{PSIi}). Finally the last structure equation and
second-Cartan curvature two-form were introduced by

\begin{eqnarray}
\Omega _{i} &\equiv &\frak{\ \ }{\large \mathfrak{D} }\Psi _{i}=
d\Psi _{i}+\eta
^{jk}\Psi _{j}\wedge \omega _{ki}  \label{THREE} \\
&=&\frac{1}{2}(\nabla ^{j}C_{ijmn}+ A^{j}C_{ijmn})\theta
^{m}\wedge \theta ^{n}.
\end{eqnarray}

The question or issue is what is the meaning of these resulting
structures?

These equations can be unified in the following fashion: we first
group together the fifteen one-forms
\begin{equation}
\omega =(\theta ^{i},\text{ }\omega _{[ij]},\text{ }A,\text{
}\Psi_{j}), \label{wab}
\end{equation}
as the ``Cartan connection'', represented by the 6x6 matrix of
one-forms,

\begin{equation}
\omega _{\;B}^{A}\,=\left[
\begin{array}{lll}
-A & \Psi _{i} & 0 \\
\theta ^{i} & \eta ^{ik}\omega _{[kj]} & \eta ^{ij}\Psi _{j} \\
0 & \eta _{ij}\theta ^{j} & A
\end{array}
\right] ,  \label{wAB}
\end{equation}
and the curvature two-forms, ($T^{j}$ is the vanishing torsion)
\begin{equation}
R=(T^{j}=0,\Omega _{\,\,\,\,j}^{i},\Omega _{\,i}),  \label{Rab}
\end{equation}
as the ``Cartan curvature'', represented by the 6x6 matrix of
two-forms,
\begin{equation}
R_{\;B}^{A}\,=\left[
\begin{array}{lll}
0 & \Omega _{\,i} & 0 \\
0 & \Omega _{\,\,\,\,j}^{i} & \eta ^{ij}\Omega _{\,j} \\
0 & 0 & 0
\end{array}
\right] .  \label{RAB}
\end{equation}
One can then see, by a straightforward calculation,
that\textit{remarkably} the three structure equations
Eqs.(\ref{ONE}), (\ref{TWO}) and (\ref{THREE}) are all encompassed
in the single Cartan structure equation

\begin{equation}
R_{\;B}^{A}= d\omega _{\;B}^{A}+\omega _{\;C}^{A}\wedge \omega
_{\;B}^{C}. \label{CartanStEq}
\end{equation}
One finds that the one and two-form matrices $\omega _{\;B}^{A}\,\
$and $R_{\;B}^{A}$ take their values in the Lie algebra of the
fifteen parameter group $O(4,2)$\cite{Kob}, though as forms they
are in the six-space $J^{6}$. The $O(4,2)$ Lie algebra is graded
as
\[
O(4,2)^{\prime }\frak{=g}_{-1}+\frak{g}_{0}+\frak{g}_{1}
\]
with
\begin{eqnarray}
&&\text{ \qquad } \fbox{$\theta ^{j}$ $\epsilon $
$\frak{g}_{-1}$},\emph{g}
\label{lie} \\
&&\fbox{($A$, $\omega _{\,\,\,i}^{k}$) \& $\Omega
_{\,\,\,\,i}^{k}$ $\epsilon $ $\frak{g}_{0}$},  \nonumber \\
\quad &&\quad \fbox{$\Psi _{i}$ \& $\Omega _{\,i}$ $\epsilon $
$\frak{g}_{1} $}.  \nonumber
\end{eqnarray}

Aside from the fact that the dimension count is not correct, i.e.,
the fibers are two-dimensional and are not the needed eleven
dimensions, we have \textit{all the conditions} for a Cartan
$O(4,2)$ normal conformal connection \cite{Kob}. In addition to
the correct Lie algebra, we have: the three structure equations,
Eqs.(\ref{ONE}), (\ref{TWO}) and (\ref{THREE}), zero torsion, a
trace-free first Cartan curvature $\Omega _{ij}$ (the Weyl tensor)
and a second Cartan curvature $\Omega _{i}$ with the correct
structure, i.e., vanishing fiber parts. It is clear that we are
dealing with a six-dimensional subbundle of the full
15-dimensional bundle. The fibers should be coordinatized by the
11-dimensional subgroup $H=CO(3,1)\otimes _{s}T^{*4}$ of $O(4,2).$
The question is where and what are the missing coordinates needed
to describe $H$?

The eleven coordinates (or parameters) must be such that when
acting on the conformal metric, Eq.(\ref{g}), it is left
conformally unchanged. Actually we already have seven parameters,
namely ($s$,$s^{*},\varpi ,A_{i}$); we have from Eq.(\ref{Lg=0}),
that variations in $s$ and $s^{*}$ or multiplication by $\varpi $
leaves Eq.(\ref{g}) conformally unchanged. The $A_{i}$ have no
relationship with the metric.

The other four parameters could be chosen as follows:

a) Rescaling $\theta ^{+}$ and $\theta ^{-}$ respectively, by
e$^{i\psi }$ and e$^{-i\psi }$ leaves Eq.(\ref{g}) unchanged. (The
parameters ($s,s^{*},\psi $) describe $O(3)$ transformations.)

b) One could take three-parameter, ($\gamma ,\gamma ^{*},\mu $),
linear combinations of the four $\theta ^{i}$ that constitute
Lorentz transformations with no change in the metric. For example,
we could have the ($\theta ^{0},$ $\theta ^{1})$ boost,
\begin{equation}
\theta ^{^{\prime }0}= \mu \theta ^{0},\text{ }\theta ^{^{\prime
}1}=\mu ^{-1}\theta ^{1},
\end{equation}
and the null rotations given by,

\begin{eqnarray}
\theta ^{^{\prime }0} &=&\theta ^{0}, \\
\theta ^{^{\prime }+} &=&\theta ^{+}+\gamma \theta ^{0}, \\
\theta ^{^{\prime }-} &=&\theta ^{-}+\gamma ^{*}\theta ^{0}, \\
\theta ^{^{\prime }1} &=& \theta ^{1}+\gamma \theta ^{-}+\gamma
^{*}\theta ^{+}+\gamma \gamma ^{*}\theta ^{0+}.
\end{eqnarray}

The seven $(s,s^{*},\psi ,\gamma ,\gamma ^{*},\mu ,\varpi $)
parametrize the conformal Lorentz group, $CO(3,1)$ while $A_{i}$
parametrize $T^{*4}.$ With the exception of $(s,s^{*})$ all the
remaining parameters, $(\psi ,\gamma ,\gamma ^{*},\mu ,\varpi
,A_{i}$), are chosen as arbitrary functions on $ \frak{M}^{4}.$

It would have been possible to start with a generalized version of
the $ \theta ^{i}$ [see remark \#2] so that all eleven parameters
in our equations appeared at the start. In the present work, we
have effectively taken a six-dimensional subbundle
(two-dimensional fibers) by choosing $\gamma $ $=$ $\gamma
^{*}=\psi =0,$ $\mu =1,$ with $\varpi $ an arbitrary but given
function on $\frak{M}^{4}$ and $A_{i}$ four arbitrary functions on
$J^{6}$. Since the $A_{i}$ are arbitrary it is possible (and
probably more attractive) to choose them to also be functions just
on $\frak{M}^{4}.$ Only the ($s,s^{*})$ can vary on each fiber.

\section{Conclusion}

The work presented here addresses the issue of how
four-dimensional differential geometry can be coded into pairs of
2$^{nd}$ order PDE's. More specifically, we have shown how all
Cartan $O(4,2)$ normal conformal connections can be so coded. Our
ultimate goal, however, is a step beyond this; we want to know how
to code the conformal Einstein equations into such pairs of
2$^{nd}$ order PDE's. This would mean further conditions, over and
above the W\"{u}nschmann condition, on the choice of the $S$ and
$S^{*}.$ Though at the present time we do not know the details of
these `further conditions'. Nevertheless, there is a clear
strategy for their determination.

It appears that they can be expressed as the vanishing of two (or
three) different functionals of
$S(Z,Z_{s},Z_{s^{*}},Z_{ss^{*}},s,s^{*})$ and
$S^{*}(Z,Z_{s},Z_{s^{*}},Z_{ss^{*}},s,s^{*})$. These functionals
can be derived from the vanishing of the Bach tensor\cite{Lew} and
an algebraic restriction on the Cartan curvature
$R_{B}^{A}$\cite{KNN}$,$ i.e., from

\begin{eqnarray}
\nabla ^{m}\nabla ^{n}C_{mabn}+\frac{1}{2}R^{mn}C_{mabn} &=& 0,
\label{bach}
\\
C^{efgh}[C_{efgh}\nabla ^{d}C_{cdab}-4\nabla^{d}C_{efgd}C_{chab}]
&=&0. \label{cubicC}
\end{eqnarray}
It is known that both the vanishing of the Bach tensor,
Eq.(\ref{bach}), and this restriction on the Cartan curvature,
Eq.(\ref{cubicC}), yields metrics that are conformally related to
vacuum Einstein metrics. It seems very likely that in the context
of the Cartan connection these tensor equations, with their many
components, can be reduced to simply two (or three) equations for
the $S$ and $S^{*}.$

At the moment, the problem appears to be algebraically quite
formidable, however, with computer algebra available, it seems to
be manageable. Work on the problem has begun.

\section{Acknowledgments}

We happily acknowledge, with many thanks, the help and knowledge
given to us by Pawel Nurowski. ETN and KP thank the National
Science Foundation for support under Grants No. 0088951 and
0244513. CK and EG thank CONICET and the NSF for financial
assistance.

\section{\textbf{Appendices}}

In appendix A we give the relationship between the hatted and
unhatted quantities and then explicitly display the hatted
quantities in appendix B. In appendix C we find the tetrad
parameters as functions of $(S,S^{*}).$

\subsection{Appendix}

The expressions for the $\triangle ^{\prime }s$, the $G^{\prime
}$s and the $\omega ^{\prime }s$ are most easily expressed through
their hatted counterparts. The relations between the hatted and
unhatted quantities are given by:

\begin{eqnarray}
\triangle \,_{ijs} &=& \widehat{\triangle }\,_{ijs}-\eta _{ij}\Phi
^{-1}D\Phi
,  \label{triangletriangles} \\
\triangle \,_{ijk} &=&\widehat{\triangle }\,_{ijk}\Phi ^{-1}+
2e_{[j}\Phi \,\eta _{k]\,\,i}\Phi ^{-2},  \label{triangletriangle}
\end{eqnarray}
\newline

\begin{equation}
G_{ij}= \widehat{G}_{ij}+2\eta _{ij}\Phi ^{-1}D\Phi ,
\label{GGhat2}
\end{equation}
\begin{equation}
\omega _{ijk}= \widehat{\omega }_{ijk}\Phi
^{-1}+2\widehat{e}_{[j}\Phi \cdot \eta _{i]k\,\,}\Phi ^{-2},
\end{equation}

\begin{eqnarray}
\omega _{ijs} &=&\widehat{\omega }_{ijs}-\eta _{ij}\Phi ^{-1}D\Phi ,
\label{wwhat2} \\
\omega _{ijs^{*}} &=&\widehat{\omega }_{ijs^{*}}-\eta _{ij}\Phi
^{-1}D^{*}\Phi ,  \nonumber
\end{eqnarray}

\begin{eqnarray}
\widehat{A}_{s} &=&
-\frac{1}{2}S_{W}+\frac{b^{*}S_{W^{*}}-2ab}{(1+bb^{*})} +
\frac{a^{*}(1+6bb^{*}+b^{2}b^{*2})}{2(1+bb^{*})^{2}}
\Rightarrow A_{s}=0. \\
A_{i} &=&\widehat{A}_{i}\Phi ^{-1}
\end{eqnarray}

They are easily derived from their definitions, (Secs. II and III)
using the two different choices of $\Phi ,$ $(\Phi =1$ and $\Phi
=\varpi \Phi _{0}).$ Note that from the differential equation for
$\Phi ,$ Eq.(\ref{Dphi}), we have that
\[
\Phi ^{-1}D\Phi =\frac{1}{4}\widehat{\triangle }^{k}\,_{ks}.
\]

In the next appendix the hatted quantities ($\widehat{\triangle
}$, $\widehat{G}$ , $\widehat{\omega })$ are explicitly displayed
as functions of ($S$, $S^{*},$ $\widehat{A}_{i}$), leading to the
expressions for ($\triangle ,$ $G$, $\omega )$ in terms of $(S$,
$S^{*},$ $\widehat{A}_{i},\varpi $).

\subsection{\textbf{Appendix }}

Defining the quantities

\begin{eqnarray}
\gamma &\equiv &1-bb^{*}, \\
\sigma &\equiv &a-b^{*}a^{*}, \\
\zeta &\equiv &a_{R}-b^{*}a_{R}^{*}.
\end{eqnarray}
and
\begin{equation}
\hat{h}_{+}=\hat{e}_{+}+b^{*}\hat{e}_{-},
\end{equation}
(noting that $\widehat{\triangle }^{-}\,_{JK}$ and
$\widehat{G}_{ij}^{*}$ can be obtained by complex conjugation) we
have the three sets, ($\widehat{\triangle }$, $\widehat{G}$ ,
$\widehat{\omega })$:

I. The $\widehat{\triangle }$ 's:

\begin{eqnarray}
\widehat{\triangle }^{0}\,_{0+}\ \, &=&\widehat{\triangle
}^{0}\,_{0-}= \widehat{\triangle }^{0}\,_{01}\,=
\widehat{\triangle
}^{0}\,_{0s} = \widehat{\triangle }^{0}\,_{0s^{*}} =
\widehat{\triangle }^{0}\,_{+-}= 0, \\
\widehat{\triangle }^{0}\,_{+1} &=&\widehat{\triangle
}^{0}\,_{-1}= \widehat{ \triangle }^{0}\,_{1s}=\widehat{\triangle
}^{0}\,_{1s^{*}}=\widehat{
\triangle }^{0}\,_{ss^{*}}=0,  \nonumber \\
\widehat{\triangle }^{0}\,_{+s}\;\, &=&\widehat{\triangle
}^{0}\,_{-s^{*}}=
\frac{-1}{\alpha \gamma },  \nonumber \\
\widehat{\triangle }^{0}\,_{+s^{*}} &=&\frac{b^{*}}{\alpha \gamma },
\nonumber \\
\widehat{\triangle }^{0}\,_{-s}\; &=& \frac{b}{\alpha \gamma },
 \nonumber \\
\widehat{\triangle }^{+}\,_{0+}\; &=&\hat{e}_{0}(\ln \alpha
)-\frac{b_{0}^{*}
\hat{e}(b)}{\gamma },  \nonumber \\
\widehat{\triangle }^{+}\,_{0-}\; &=&\frac{\hat{e}_{0}(b)}{\gamma },
\nonumber \\
\widehat{\triangle }^{+}\,_{01}\; &=&0,  \nonumber \\
\widehat{\triangle }^{+}\,_{0s}\;\; &=&\alpha (bc-\hat{e}_{0}(S)),
\nonumber
\\
\widehat{\triangle }^{+}{}_{0s^{*}} &=& \alpha
(c-b\hat{e}_{0}(S^{*})),
\nonumber \\
\widehat{\triangle }^{+}\,_{+-}\, &=& \frac{\hat{h}_{+}(b)}{\gamma
}-\hat{e}
_{-}(\ln \alpha ),  \nonumber \\
\widehat{\triangle }^{+}\,_{+1}\; &=& \frac{\alpha
b^{*}b_{R}-\gamma \alpha
_{R}}{\alpha \gamma },  \nonumber \\
\widehat{\triangle }^{+}\,_{+s}\; &=& -D(\ln \alpha
)+\frac{b^{*}Db-\alpha
\gamma \hat{e}_{+}(S)+b\sigma }{\gamma },  \nonumber \\
\widehat{\triangle }^{+}\,_{+s^{*}} &=&-D^{*}(\ln \alpha )+\frac{
b^{*}D^{*}b-\alpha \gamma b\hat{e}_{+}(S^{*})+\sigma }{\gamma },
\nonumber
\\
\widehat{\triangle }^{+}\,_{-1}\; &=&\frac{-b_{R}}{\gamma },
\nonumber \\
\widehat{\triangle }^{+}\,_{-s}\; &=&\frac{-Db-\alpha \gamma
\hat{e}
_{-}(S)+b\sigma ^{*}}{\gamma },  \nonumber \\
\widehat{\triangle }^{+}\,_{-s^{*}} &=& \frac{-D^{*}b-\alpha
\gamma b\hat{e}
_{-}(S^{*})+\sigma ^{*}}{\gamma },  \nonumber \\
\widehat{\triangle }^{+}\,_{1s}\;\; &=&-\alpha (b+S_{R}),
\nonumber \\
\widehat{\triangle }^{+}\,_{1s^{*}} &=&-\alpha (1+bS_{R}^{*}),
\nonumber \\
\widehat{\triangle }^{+}\,_{ss^{*}} &=&0,  \nonumber
\end{eqnarray}

\begin{eqnarray}
\widehat{\triangle }^{1}\,_{0+}\; &=&-\hat{e}_{+}(c)+\frac{\hat{e}
_{0}(a)-b_{0}^{*}\hat{e}(a^{*})}{\alpha \gamma }, \\
\widehat{\triangle }^{1}\,_{0-}\; &=&-\hat{e}_{-}(c)+\frac{\hat{e}
_{0}(a^{*})-b\hat{e}_{0}(a)}{\alpha \gamma },  \nonumber \\
\widehat{\triangle }^{1}\,_{01}\; &=&-c_{R},  \nonumber \\
\widehat{\triangle }^{1}\,_{0s}\;\;
&=&-Dc+a^{*}c-\hat{e}_{0}(T)-a\hat{e}
_{0}(S),  \nonumber \\
\widehat{\triangle }^{1}\,_{0s^{*}} &=&-D^{*}c+ac-\hat{e}
_{0}(T^{*})-a_{0}^{*}\hat{e}(S^{*}),  \nonumber \\
\widehat{\triangle }^{1}\,_{+-}\,
&=&\frac{\hat{h}_{+}(a^{*})-\hat{h}_{-}(a)
}{\alpha \gamma },  \nonumber \\
\widehat{\triangle }^{1}\,_{+1}\ &=&\frac{-\zeta }{\alpha \gamma },
\nonumber \\
\widehat{\triangle }^{1}\,_{+s}\ &=&
-(\hat{e}_{+}(T)+a\hat{e}_{+}(S))+\frac{b^{*}Da^{*}-Da-c+a^{*}\sigma
}{\alpha \gamma },
\nonumber \\
\widehat{\triangle }^{1}\,_{+s^{*}} &=&
-(\hat{e}_{+}(T^{*})+a_{+}^{*}\hat{e}
(S^{*}))+\frac{b^{*}(D^{*}a^{*}+c)-D^{*}a+a\sigma }{\alpha \gamma
},
\nonumber \\
\widehat{\triangle }^{1}\,_{-1}\; &=& \frac{-\zeta ^{*}}{\alpha
\gamma },
\nonumber \\
\widehat{\triangle }^{1}\,_{-s}\;
&=& -(\hat{e}_{-}(T)+a\hat{e}_{-}(S))+
\frac{b(Da+c)-Da^{*}+a^{*}\sigma ^{*}}{\alpha \gamma },
 \nonumber \\
\widehat{\triangle }^{1}\,_{-s^{*}} &=&
-(\hat{e}_{-}(T^{*})+a_{-}^{*}\hat{e}
(S^{*}))+\frac{bD^{*}a-D^{*}a^{*}-c+a\sigma ^{*}}{\alpha \gamma },
\nonumber
\\
\widehat{\triangle }^{1}\,_{1s}\;\; &=& -(T_{R}+a^{*}+aS_{R}),
\nonumber \\
\widehat{\triangle }^{1}\,_{1s^{*}} &=&
-(T_{R}^{*}+a+a^{*}S_{R}^{*}),
\nonumber \\
\widehat{\triangle }^{1}\,_{ss^{*}} &:&=0.  \nonumber
\end{eqnarray}

II. The $\widehat{G}$ 's:

\begin{eqnarray}
\widehat{G}_{00}\,\, &=& 2(Dc + aS_{Z} + T_{Z}- c(aS_{R} +
T_{R} + a^{*})), \\
\widehat{G}_{0+}\, &=& \alpha ^{-1}(1-bb^{*})^{-1}\{\alpha
^{2}(1-bb^{*})[c(1+b^{*}S_{R})-b^{*}S_{Z}]  \nonumber \\
&&-b^{*}(Da^{*}-
aa^{*}S_{R}+aS_{W^{*}}-(a^{*})^{2}+T_{W^{*}}-a^{*}T_{R})
\nonumber \\
&&+Da+c+T_{W}+aS_{W}-aT_{R}-a^{2}S_{R}-aa^{*}\},  \nonumber \\
\widehat{G}_{0-} &=&\alpha ^{-1}(1-bb^{*})^{-1}\{\alpha
^{2}(1-bb^{*})[c(b+S_{R})-S_{Z}]  \nonumber \\
&&-b(c+Da-aT_{R}-aa^{*}-a^{2}S_{R}+aS_{W}+T_{W})  \nonumber \\
&&+Da^{*}+T_{W^{*}}+
aS_{W^{*}}-a^{*}T_{R}-aa^{*}S_{R}-(a^{*})^{2}\},
\nonumber \\
\widehat{G}_{01}\,\, &=&a^{*}+aS_{R}+T_{R},  \nonumber \\
\widehat{G}_{++} &=&-2\;(1-bb^{*})^{-1}
\{b^{*}(a^{*}-b^{*}S_{W^{*}}+a^{*}b^{*}S_{R}-aS_{R}+S_{W})+
Db^{*}-a\},
\nonumber \\
\widehat{G}_{+-} &=& \alpha
^{-1}(1-bb^{*})^{-1}\{(1-bb^{*})[\alpha
(a^{*}-S_{W}+aS_{R})-2D\alpha ] + \alpha D(bb^{*})\},
\nonumber \\
\widehat{G}_{+1} &=&\alpha ^{-1}(1-bb^{*})^{-1}\{1-\alpha
^{2}(1-bb^{*})(1+b^{*}S_{R})\},  \nonumber \\
\widehat{G}_{--}
&=&-2\;(1-bb^{*})^{-1}\{b(ab-a^{*}+aS_{R}-S_{W})+Db-a^{*}S_{R}+
S_{W^{*}}\},
\nonumber \\
\widehat{G}_{-1}\, &=&\alpha ^{-1}(1-bb^{*})^{-1}\{-b-\alpha
^{2}(1-bb^{*})(b+S_{R})\},  \nonumber \\
\widehat{G}_{11}\,\, &=&0.  \nonumber
\end{eqnarray}
III. The $\widehat{\omega }$ 's:

\begin{eqnarray}
\widehat{\omega }_{0+}
&=&\{\hat{e}_{+}(c)+\frac{b_{0}^{*}\hat{e}(a^{*})-
\hat{e}_{0}(a)}{\alpha \gamma }\}\widehat{\theta
}^{0}+\{\widehat{A}_{+}- \frac{\zeta }{2\alpha \gamma
}\}\widehat{\theta }^{1}+\frac{\hat{e}
_{0}(b^{*})}{\gamma }\widehat{\theta }^{+} \\
&&+\{\widehat{A}_{0}+\frac{2\hat{e}_{0}(bb^{*})+\alpha \gamma
^{2}(\hat{h} _{+}(a^{*})-\hat{h}_{-}(a))}{2\gamma
(1+bb^{*})}\}\widehat{\theta }^{-}
\nonumber \\
&&+\frac{\gamma c-b^{*}S_{Z}(1+bb^{*})}{\alpha \gamma
^{2}}\widehat{\theta } ^{s}-\frac{b^{*}\gamma
c+S_{Z}^{*}(1+bb^{*})}{\alpha \gamma ^{2}}\widehat{ \theta
}^{s^{*}},  \nonumber
\end{eqnarray}

\[
\widehat{\omega }_{0-}=(\widehat{\omega }_{0+})^{*},
\]
\begin{eqnarray*}
\widehat{\omega }_{01} &=&c_{R}^{0}\widehat{\theta
}^{0}+\{\widehat{A}_{+}+ \frac{\zeta }{2\alpha \gamma
}\}\widehat{\theta }^{+}+\{\widehat{A}_{-}+ \frac{\zeta
^{*}}{2\alpha \gamma }\}\theta ^{-}+2\widehat{A}_{1}\widehat{
\theta }^{1} \\
&&+2\widehat{A}_{s}\widehat{\theta }^{s}+
2\widehat{A}_{s^{*}}\widehat{\theta }^{s^{*}},
\end{eqnarray*}
\[
\widehat{\omega }_{10}=-\widehat{\omega }_{01}+2\widehat{A},
\]

\begin{eqnarray*}
\widehat{\omega }_{+-}
&=&\{-A_{0}+\frac{b_{0}^{*}\hat{e}(b)-b\hat{e}
_{0}(b^{*})}{2\gamma }+\frac{\alpha \gamma (\hat{h}_{-}(a)-\hat{h}
_{+}(a^{*}))}{2(1+bb^{*})}\}\widehat{\theta }^{0} \\
&&-\{\widehat{A}_{1}+\frac{bb_{R}^{*}-b^{*}b_{R}}{2\gamma
}\}\widehat{\theta }^{1}+\{-\frac{\hat{h}_{-}(b^{*})}{\gamma
}+\frac{(3+bb^{*})\hat{e}
_{+}(bb^{*})}{2\gamma (1+bb^{*})}\}\widehat{\theta }^{+} \\
&&-\{-\frac{\hat{h}_{+}(b)}{\gamma
}+\frac{(3+bb^{*})\hat{e}_{-}(bb^{*})}{
2\gamma (1+bb^{*})}+2\widehat{A}_{-}\}\widehat{\theta }^{-} \\
&&-\{\frac{\gamma
(S_{W}+2A_{s})+a^{*}(3+bb^{*})}{4}-\frac{ab(1+3bb^{*})}{
2(1+bb^{*})}\}\widehat{\theta }^{s} \\
&&+\{\frac{\gamma
S_{W^{*}}^{*}+(a-2A_{s^{*}})(3+bb^{*})}{4}-\frac{
a^{*}b^{*}(1+3bb^{*})}{2(1+bb^{*})}\}\widehat{\theta }^{s^{*}},
\end{eqnarray*}
\[
\widehat{\omega }_{-+}=-\widehat{\omega }_{+-}-2\widehat{A},
\]

\begin{eqnarray*}
\widehat{\omega }_{+1} &=& \{-\widehat{A}_{+}+\frac{\zeta
}{2\alpha \gamma }\} \widehat{\theta }^{0}-\frac{b_{R}^{*}}
{\gamma }\widehat{\theta }^{+}-\{
\widehat{A}_{1}+\frac{(bb^{*})_{R}}{\gamma
(1+bb^{*})}\}\widehat{\theta }^{-}
\\
&&\quad +\frac{\alpha \gamma }{1+bb^{*}}\widehat{\theta }^{s}-
\frac{\alpha \gamma b^{*}}{1+bb^{*}}\widehat{\theta }^{s^{*}},
\end{eqnarray*}

\[
\widehat{\omega }_{-1}=(\widehat{\omega }_{+1})^{*}.
\]

\subsection{\textbf{Appendix}}

\qquad In this appendix, which is long and very complicated but
given for completeness, we obtain the tetrad parameters and
conditions on $S$ and $ S^{*}$ that uniquely determine our torsion
free connection. The vanishing of the trace free part of
$\widehat{\triangle }\,_{(ij)s}$, i.e., conditions (
\ref{GhatTF}), gives

\begin{eqnarray}
\widehat{G}_{01}+\widehat{G}_{+-}
&=&\widehat{G}_{01}^{*}+\widehat{G}
_{+-}^{*}=0, \\
\widehat{G}_{ij} &=&\widehat{G}_{ij}^{*}= 0,\text{ for
}(i,j)\,\not {\in}\,\{(0,1),(+,-)\}.
\end{eqnarray}

The explicit expressions for $\widehat{G}_{ij}$ given in the
previous appendix are used to \emph{i)} solve for the tetrad
parameters and \emph{ii)} derive the W\"{u}nschmann condition.

\emph{In what follows, we will often encounter pairs of equations
that are complex-conjugate to one another. In these instances, we
will list only one of the equations and imply the other. When we
want to refer to the conjugate of a listed equation, we will
write the listed equation's reference number with a super-script
(*).}

We start with the equations $\widehat{G}_{+1}=0$,
$\widehat{G}_{-1}=0$, $ \widehat{G}_{+1}^{*}=0$, and
$\widehat{G}_{-1}^{*}=0$, which depend only on $ b$, $b^{*}$, and
$\alpha $. They are four equations for three variables that
satisfy an identity. From $\widehat{G}_{+1}=0$ and
$\widehat{G}_{-1}^{*}=0$, we have
\begin{equation}
b^{*}S_{R}=bS_{R}^{*}.  \label{bandbstar}
\end{equation}
Next, using $\widehat{G}_{-1}^{*}=0$ and $\widehat{G}_{-1}=0$ to
eliminate $ \alpha ^{2}$, we obtain
\begin{equation}
b=\frac{-1+\sqrt{1-S_{R}S_{R}^{*}}}{S_{R}^{*}}.  \label{eqb}
\end{equation}
(We have chosen the positive root since we want $b$ to vanish when
$S$ vanishes.) One sees that $b^{*}$ is the complex conjugate of
$b$. It is useful to invert Eqs.(\ref{eqb}) and (\ref{eqb}$^{*}$),
yielding
\begin{eqnarray}
S_{R} &=&\frac{-2b}{1+bb^{*}}\text{ ,}  \label{eqsr} \\
\text{ }S_{R}^{*} &=&\frac{-2b^{*}}{1+bb^{*}}.  \nonumber
\end{eqnarray}
From Eq.(\ref{eqsr}) and $\widehat{G}_{+1}=0$, we find
\begin{equation}
\alpha ^{2}=\frac{1+bb^{*}}{(1-bb^{*})^{2}}.  \label{eqalphasq}
\end{equation}
All four equations, \{$\widehat{G}_{+1}=0$, $\widehat{G}_{-1}=0$,
$\widehat{G }_{+1}^{*}=0$, $\widehat{G}_{-1}^{*}=0$\}, are
satisfied by Eqs.(\ref{eqb}), (\ref{eqb}$^{*}$) and
(\ref{eqalphasq}).

Our next step is to determine $a$, $a^{*}$, and the W\"{u}nschmann
condition from the equations
$\widehat{G}_{01}+\widehat{G}_{+-}=0$, $\widehat{G} _{++}=0 $,
$\widehat{G}_{--}=0$, and their conjugates. From this set of six
equations we will be able to solve for ($a$, $a^{*}),$ find
restrictions on ( $S,S^{*})$ and obtain further identities.

We first state some useful relationships. Taking $D$ of
Eq.(\ref{eqalphasq}) we have, after some simplification,
\begin{equation}
D\alpha =\frac{\alpha D(bb^{*})(3+bb^{*})}{2(1+bb^{*})(1-bb^{*})}.
\label{eqdalpha}
\end{equation}
Next, (see Eq.(\ref{t})) we find $T_{R}=T_{R}[Db,Db^{*},b,S]$ and
$ T_{R}^{*}=T_{R}^{*}[Db,Db^{*},b,S]$. By first taking $D^{*}$ of
Eq.(\ref {eqsr}) and $D$ of Eq.(\ref{eqsr}$^{*}$), i.e.,
\begin{eqnarray}
D^{*}(S_{R}) &=&D^{*}(\frac{-2b}{1+bb^{*}}),  \label{DSr} \\
D^{*}(S_{R}^{*}) &=&D(\frac{-2b^{*}}{1+bb^{*}}),  \nonumber
\end{eqnarray}
then using, Eq.(\ref{comm}), to commute the $R$-derivative and the
fiber-derivative, we obtain two equations containing $T_{R}$ and
$T_{R}^{*}$ . After simplifying with Eqs.(\ref{eqsr}), they become

\begin{eqnarray}
T_{R} &=&\frac{4b(Db^{*}-b^{*2}Db)}{(1+bb^{*})(1-bb^{*})^{2}}+
\frac{ 2(b^{2}D^{*}b^{*}-D^{*}b+2b^{2}S_{W}^{*})}{(1-bb^{*})^{2}}
\label{Tr} \\
&&+\frac{S_{W}(1+bb^{*})^{2}}{(1-bb^{*})^{2}}-\frac{
2(1+bb^{*})(b^{*}S_{W^{*}}+bS_{W^{*}}^{*})}{(1-bb^{*})^{2}}.
\nonumber
\end{eqnarray}

We are now in a position to find $a,$ $a^{*}$, and the
W\"{u}nschmann condition. First, from
$\widehat{G}_{01}+\widehat{G}_{+-}=0$ and $\widehat{G}
_{01}^{*}+\widehat{G}_{+-}^{*}=0,$ we solve for $a$ and $a^{*}$.
With the aid of Eqs.(\ref{eqdalpha}), (\ref{Tr}) and their
conjugates, we find
\begin{equation}
a=\frac{ (1+bb^{*})[b^{*2}Db+
Db^{*}+D^{*}(bb^{*})+(1-bb^{*})(b^{*}S_{W}+bS_{W}^{*})]}{
(1-bb^{*})^{3}}.  \label{eqfirsta}
\end{equation}
When they are inserted into $\widehat{G}_{--}=0$, we find that
$S$ must obey the differential condition
\begin{equation}
M \equiv \frac{Db+bD^{*}b+S_{W^{*}}-bS_{W}+b^{2}S_{W^{*}}^{*}-
b^{3}S_{W}^{*}}{ 1-bb^{*}}=0,  \label{eqm}
\end{equation}
where $b$ is the known expression in terms of $S$ and $S^{*}$. The
expression $M=M[Db,D^{*}b,b]$ is known as the generalized
W\"{u}nschmann invariant. Its vanishing is the condition on the
$S$ and $S^{*},$ i.e., on the original pair of PDE's, for the
existence of a torsion-free connection.

This condition tells us that this invariant must vanish if we are
to find a non-trivial torsion free connection. By substituting
$Db^{*},$ from the W\"{u}nschmann invariant and its conjugate,
into Eqs.(\ref{eqfirsta}) and (\ref{eqfirsta}$^{*}$), our
expression for $a$ becomes

\begin{eqnarray}
a &=& b^{-1}b^{*-1}(1-bb^{*})^{-2}(1+bb^{*})\{b^{*2}(M-Db+bS_{W}-
S_{W^{*}})
\label{eqa} \\
&&\quad +b(M^{*}-D^{*}b^{*}+b^{*}S_{W^{*}}^{*}-S_{W}^{*})\}.
\nonumber
\end{eqnarray}
or, with $M=M^{*}=0,$ we have

\begin{eqnarray}
a &=&b^{-1}b^{*-1}(1-bb^{*})^{-2}(1+bb^{*})\{b^{*2}(bS_{W}-Db-
S_{W^{*}})
\label{a2} \\
&&\quad +b(b^{*}S_{W^{*}}^{*}-D^{*}b^{*}-S_{W}^{*})\}.  \nonumber
\end{eqnarray}

Summarizing our results so far; we have obtained the five tetrad
parameters, ($b,b^{*},\alpha ,a,a^{*}$), as well as the
W\"{u}nschmann condition in terms of $S$ and $S^{*}$. The search
for the last parameter, i.e., $c$, is the most interesting and at
the same time the most difficult part of the construction.

There are four equations, namely, $\widehat{G}_{0+}=0$,
$\widehat{G}_{0-}=0$ , $\widehat{G}_{0+}^{*}=0$, and
$\widehat{G}_{0-}^{*}=0$, for $c$. As we will see below, three of
those equations become identities once we algebraically solve for
$c$. It is, however, instructive to keep the W\"{u}nschmann
invariant different from zero when solving the equations. We then
explicitly show how its vanishing yields a unique solution for
$c,$ as well satisfying the remaining identities among the
$\widehat{G}_{ij}$'s. Thus, for the subsequent calculations, $M$
is left in the equations. We begin by
\begin{eqnarray}
Db &=& M+bS_{W}-S_{W^{*}}+\frac{b(1-bb^{*})(ab-a^{*})}{1+bb^{*}},
\label{eqdb} \\
Db^{*} &=&b^{*}(b^{*}S_{W^{*}}-S_{W})-bM^{*}+
\frac{(1-bb^{*})(a-a^{*}b^{*})}{ 1+bb^{*}}.  \label{eqdbstar}
\end{eqnarray}

Next, we insert the left-hand sides of Eqs. (\ref{eqdb}),
(\ref{eqdbstar}), and their conjugates into Eqs. (\ref{Tr}) and
(\ref{Tr}$^{*}$) to find $ T_{R}=T_{R}[a,b;M]$ and
$T_{R}^{*}=T_{R}^{*}[a,b;M]$. The result is
\begin{equation}
T_{R}=(\tau +S_{W})+\frac{2(3ab-b^{*}S_{W^{*}})}{1+bb^{*}}-\frac{
2a^{*}(1+4bb^{*}+b^{2}b^{*2})}{(1+bb^{*})^{2}},  \label{eqtr}
\end{equation}
where $\tau = \tau [M]$ (which vanishes with $M$) is given
below\textbf{.}

Third, using the integrability condition, one shows that the
vectors $D$ and $D^{*}$ commute. In particular,

\begin{eqnarray}
DD^{*}b &=&D^{*}Db,  \label{commddstarb} \\
DD^{*}b^{*} &=&D^{*}Db^{*}.  \nonumber
\end{eqnarray}
Thus by taking the appropriate fiber-derivatives of the four
Eqs.(\ref{eqdb} ), (\ref{eqdb}$^{*}$), (\ref{eqdbstar}), and
(\ref{eqdbstar}$^{*}$), simplifying with Eqs.(\ref{comm}),
(\ref{eqdb}), (\ref{eqdbstar}), and their conjugates, and by using
the Eqs.(\ref{commddstarb}), we obtain two equations containing
$Da$, $D^{*}a$, $Da^{*}$, and $D^{*}a^{*}$. They can be solved for
$Da^{*}=Da^{*}[Da,D^{*}a^{*}]$ and $D^{*}a=D^{*}a[Da,D^{*}a^{*}]$
to find

\begin{eqnarray}
Da^{*} &=& (\Upsilon +a^{*}S_{W}-a^{*2}-T_{W^{*}})+\frac{
S_{Z}(1+b^{2}b^{*2})+2b^{2}S_{Z}^{*}}{(1-bb^{*})^{2}}
\label{firstdastar} \\
&&+\frac{b(Da-D^{*}a^{*}+T_{W}-T_{W^{*}}^{*}+4aa^{*})-
2a^{*}b^{*}S_{W^{*}}}{
(1+bb^{*})}  \nonumber \\
&&
-\frac{aS_{W^{*}}(1+b^{2}b^{*2})+2b^{2}(2a^{2}+a^{*}S_{W}^{*})}{
(1+bb^{*})^{2}}.  \nonumber
\end{eqnarray}
The term $\Upsilon =\Upsilon [DM,D^{*}M,M],$ which vanishes with
$M,$ is given below. Finally, in addition to
Eq.(\ref{firstdastar}) above, we can use the integrability
condition to derive another identity on the fiber-derivatives of
$a$ and $a^{*}$. We begin by taking $D^{*}$ of Eq.(\ref{eqtr}):
\begin{equation}
D^{*}(T_{R})=D^{*}[(\tau
+S_{W})+\frac{2(3ab-b^{*}S_{W^{*}})}{1+bb^{*}}-
\frac{2a^{*}(1+4bb^{*}+b^{2}b^{*2})}{(1+bb^{*})^{2}}].
\end{equation}
On the left-hand side, we use Eq.(\ref{comm}$^{*}$) to commute the
$R$ -derivative and the fiber-derivative so that we obtain the
term $ U_{R}=\partial _{R}(D^{*}T)$, where

\begin{equation}
U\equiv D^{*}T=D^{*2}S=D^{2}S^{*}=DT^{*}
\end{equation}
denotes the integrability condition. We can then solve this
equation for $ U_{R}$. Refer to the $U_{R}$ that we obtain in this
manner as $U_{R}^{(1)}$. In a similar manner, we can obtain
$U_{R}^{(2)}$ by taking $D$ of Eq.(\ref {eqtr}$^{*}$). Then equate
$U_{R}^{(1)}$ and $U_{R}^{(2)}$, from which we find an identity on
$Da$ and $D^{*}a^{*}$. With the use of Eqs.(\ref{comm}),
(\ref{eqsr}), (\ref{eqdb}), (\ref{eqdbstar}), (\ref{firstdastar}),
and their conjugates this identity becomes

\begin{eqnarray}
&& (\Gamma -\Gamma ^{*}+Da-D^{*}a^{*}+T_{W}-T_{W^{*}}^{*})+\frac{
2(1+bb^{*})(bS_{Z}^{*}-b^{*}S_{Z})}{(1-bb^{*})^{2}}
\label{dadstarastar} \\
&& +\frac{4(a^{*2}b^{*}-a^{2}b)+2(ab^{*}S_{W^{*}}-
a^{*}bS_{W}^{*})}{1+bb^{*}} =0.  \nonumber
\end{eqnarray}
The terms $\Gamma =\Gamma [DM,D^{*}M,M]$ and its conjugate vanish
with $M$ and are given below. We are now in a position to find $c$
from the four equations $\widehat{G}_{0+}=0$,
$\widehat{G}_{0-}=0$, $\widehat{G} _{0+}^{*}=0 $, and
$\widehat{G}_{0-}^{*}=0$ . We algebraically solve each of the four
equations for $c,$ calling each solution $c^{(i)}.$ Next, we
replace $S_{R}$ , $\alpha ^{2}$, $T_{R}$, and $Da^{*}$ by Eqs.
(\ref{eqsr}), (\ref{eqalphasq}), (\ref{eqtr}), and
(\ref{firstdastar}), and use Eq.(\ref {dadstarastar}) to simplify.
Finally, we separate each $c^{(i)}$ into a piece that contains all
terms with the W\"{u}nschmann condition and its fiber-derivatives,
namely $\xi ^{(i)}$, and another piece that contains no
W\"{u}nschmann terms, namely $C^{(i)}$, so that $c^{(i)}$ has the
form
\[
c^{(i)}=C^{(i)}+\xi ^{(i)},
\]
for all $i$. It is straightforward to verify that the four
$C^{(i)}$ are real and equal. Imposing the W\"{u}nschmann
condition, $M=M^{*}=0$, so that the $\xi ^{(i)}$ $=0$, then
$C^{(i)}=c^{(i)}=c$, and we have our final expression for $c$,
namely

\begin{eqnarray}
c &=&-\frac{Da+D^{*}a^{*}+T_{W}+T_{W^{*}}^{*}}{4}-\frac{
aa^{*}(1+6bb^{*}+b^{2}b^{*2})}{2(1+bb^{*})^{2}}  \label{eqc} \\
&&+\frac{(1+bb^{*})(bS_{Z}^{*}+b^{*}S_{Z})}{2(1-bb^{*})^{2}}+\frac{
a(2ab-b^{*}S_{W^{*}})+a^{*}(2a^{*}b^{*}-bS_{W}^{*})}{2(1+bb^{*})}.
\nonumber
\end{eqnarray}

Had the W\"{u}nschmann invariant been nonvanishing the whole
construction obviously would have failed. Having determined all
the tetrad parameters, we still have to verify that
$\widehat{G}_{00}=0$ and $\widehat{G}_{00}^{*}=0$. By inspection,
we see that these equations contain fiber-derivatives of $c$. In
fact, by explicitly taking these fiber derivatives on $c,$ we find
that $ \widehat{G}_{00}=0$ and $\widehat{G}_{00}^{*}=0$ are
identically satisfied. We see this in the following fashion:

From Eqs.(\ref{firstdastar}), (\ref{firstdastar}$^{*}$),
(\ref{dadstarastar} ) and (\ref{eqc}), we find

\begin{eqnarray}
Da^{*} &=&
a^{*}S_{W}-aS_{W^{*}}-T_{W^{*}}+\frac{2a^{*}(2ab-b^{*}S_{W^{*}})}{
1+bb^{*}}  \label{dastar} \\
&&\quad +\frac{S_{Z}(1+bb^{*})^{2}}{(1-bb^{*})^{2}}-\frac{
a^{*2}(1+6bb^{*}+b^{2}b^{*2})}{(1+bb^{*})^{2}},  \nonumber
\end{eqnarray}

\begin{eqnarray}
Da &=&-(2c+T_{W})+\frac{2a(2ab-b^{*}S_{W^{*}})}{1+bb^{*}}
\label{Da} \\
&&+\frac{2b^{*}S_{Z}(1+bb^{*})}{(1-bb^{*})^{2}}-\frac{
aa^{*}(1+6bb^{*}+b^{2}b^{*2})}{(1+bb^{*})^{2}}.  \nonumber
\end{eqnarray}

By taking $D^{*}$ of Eq.(\ref{dastar}) and $D$ of
Eq.(\ref{Da}$^{*}$), subtracting them and using the commutability
of $D$ and $D^{*},$ we have

\begin{equation}
Dc=cS_{W}-T_{Z}-aS_{Z}+\frac{2c(2ab-b^{*}S_{W^{*}})}{(1+bb^{*})}-
\frac{ ca^{*}(1+6bb^{*}+b^{2}b^{*2})}{(1+bb^{*})^{2}}, \label{dc}
\end{equation}
which is equivalent to $\widehat{G}_{00}=0.$

Note that in the above analysis we have used the following
expressions, all of which vanish when $M=M^{*}=0:$

\begin{eqnarray*}
\tau &=&2(b^{*}\nu -b^{2}\nu ^{*}), \\
\Upsilon &=&2b\rho +2(1+bb^{*})^{-1}\{\mu (1-bb^{*})+\nu
[a^{*}b^{*}+a(1-bb^{*}-b^{2}b^{*2})] \\
&&+ b^{2}\mu ^{*}(1-bb^{*})+b^{2}\nu
^{*}[ab+a^{*}(1-bb^{*}-b^{2}b^{*2})]\},
\\
\Gamma &=&-b^{*}[4\mu +2\nu (2abb^{*}+a^{*}b^{*}+3a)],
\end{eqnarray*}

\begin{eqnarray*}
\xi _{1} &=& \xi _{4}^{*}=b\{\mu ^{*}(1-bb^{*})+b^{*}\rho
+\frac{1}{2}\nu
^{*}[a^{*}(3-2b^{2}b^{*2})-ab]\} \\
&&+\frac{1}{2}b^{*}\nu (a-a^{*}b^{*}),
\end{eqnarray*}

\begin{eqnarray*}
\xi _{2} &=& \xi _{3}^{*}=\frac{1}{b}\{\mu (1-bb^{*})+b\rho
+\frac{1}{2}\nu
[a(2+bb^{*}-2b^{2}b^{*2})-a^{*}bb^{*2}]\} \\
&&+\frac{1}{2}b\nu ^{*}(a^{*}-ab),
\end{eqnarray*}

where\quad
\begin{eqnarray*}
\mu &\equiv &2^{-1}(1-bb^{*})^{-3}(1+bb^{*})\{(b^{*}DM+D^{*}M) \\
&&+M(b^{*2}S_{W^{*}}-2b^{*}S_{W}-2bS_{W}^{*}+S_{W^{*}}^{*})\}, \\
\rho &\equiv &-2^{-1}(1-bb^{*})^{-2}(1+bb^{*})MM^{*}, \\
\nu &\equiv &(1-bb^{*})^{-1}(1+bb^{*})^{-1}M.
\end{eqnarray*}


\begin{thebibliography}{99}
\bibitem{T} M.A.Tresse, \textit{Determination des Invariantes Ponctuels de
l'Equation Differentielle du Second Ordre}\textbf{\ }$y^{\prime
\prime }=\omega (x,y,y^{\prime })$\textbf{, }Hirzel, Leipzig
(1896).

\bibitem{T1}  M.A. Tresse,
\textit{Sur les Invariants Differentiels des Groupes Continus de
Transformations}, Acta Math 18, 1 (1894).

\bibitem{L} S. Lie, \textit{Klassifikation und
Integration von gewohnlichen Differentialgleichungen zwischen
}$x$\textit{, }$y$ \textit{, die eine Gruppe von Transformationen
gestatten III, in Gesammelte Abhandlungen}, Vol. 5, Teubner,
Leipzig (1924).

\bibitem{C}  E. Cartan,
\textit{Sur les Varietes a Connection Projective}, Bull. Soc.
Math. France 52, 205 (1924); \ Oeuvres III, 1, N70, 825, Paris,
(1955).

\bibitem{C1}  E. Cartan,
\textit{Les Espaces Generalises e L'integration de Certaines
Classes d'Equations Differentielles}, C. R. Acad. Sci.,
\textbf{206 }, 1425 (1938).

\bibitem{C2}  E. Cartan, \textit{La Geometria de las Ecuaciones
Diferenciales de Tercer Orden}, Rev. Mat. Hispano-Amer.
\textbf{4}, 1, (1941).

\bibitem{C3} E. Cartan,
\textit{Sur une Classe d'Espaces de Weyl,} Ann. Sc.
Ec. Norm. Sup., 3e serie 60, 1 (1943).

\bibitem{Ch} S-S. Chern,
\textit{The Geometry of the Differential Equation} $y^{\prime
\prime \prime }= F(x,y,y^{\prime },y^{\prime \prime }),$ in \
Selected Papers, Springer-Verlag, (1978), original (1940)

\bibitem{W}  K. W\"{u}nschmann,
\textit{\"{U}ber Ber\"{u}hrungsbedingungen bei Integralkurven von
Differentialgleichungen,} Inaug. Dissert., Teubner, Leipzig
(1905).

\bibitem{KN}  C. N. Kozameh, E.T. Newman,
\textit{Theory of Light Cone Cuts of Null Infinity}, J. Math.
Phys., \textbf{24}, 2481 (1983).

\bibitem{FKN}  S. Frittelli, N. Kamran, E.T. Newman,
\textit{Differential Equations and Conformal Geometry,} Journal of
Geometry and Physics, \textbf{ \ \ 43} , 133 (2002).

\bibitem{etal}  S.Frittelli, C.Kozameh, E T.Newman, P. Nurowski,
\textit{Cartan Normal Conformal Connections from Differential
Equations}, Class. Quantum Grav. \textbf{19}, 5235 (2002).

\bibitem{FNN}  S.Frittelli, C. Kozameh, E.T. Newman,
\textit{Dynamics of Light-Cone Cuts of Null Infinity}, Phys. Rev.
D. \textbf{56}, 4729 (1997).

\bibitem{FCN3}  S. Frittelli, C. N. Kozameh, E.T. Newman, \textit{GR
via Characteristic Surfaces}, J. Math. Phys. \textbf{36}, 4984
(1995).

\bibitem{SCT} S. Frittelli, C. Kozameh, E.T. Newman,\
\textit{Differential Geometry from Differential Equations},
Communications in Mathematical Physics, \textbf{223}, p.383
(2001).

\bibitem{Kob}  S. Kobayashi, \textit{Transformation Groups in
Differential Geometry, }Springer-Verlag, Berlin, Heidelberg, New
York (1970)

\bibitem{Lew}  M. Korzynski, J. Lewandowski, \textit{The
Normal Conformal Cartan Connection and the Bach Tensor'', }Class.
Quantum Grav. \textbf{20}, No 16, p.3745-3764, (2003)

\bibitem{Merk} Merkulov S A , \textit{A Conformally Invariant
Theory of Gravitation and Electromagnetism}, Class. Quantum. Grav
\textbf{1}, p.349-354, (1984)

\bibitem{KNN}  C. Kozameh, E.T. Newman, P. Nurowski,
\textit{Conformal Einstein equations and Cartan conformal
connection}, Class. Quantum Grav. \textbf{20}, No 14, p.3029-3035,
(2003).
\end{thebibliography}
\end{document}